\documentclass[12pt,preprint]{article}
\usepackage[subpreambles=true]{standalone}
\usepackage{import}
\usepackage{amsmath}
\usepackage{amssymb}
\usepackage{calligra}
\usepackage{longtable}
\usepackage{authblk}
\usepackage{color}

\input epsf
\def\1{\'\i}
\providecommand{\keywords}[1]{\textbf{\textit{Keywords---}} #1}

\newcommand{\reference}[5]{\hangindent=2em\hangafter=1 #1. {\it #2}, {\bf #3}, #4, #5.}

\newcommand{\referencelink}[1]{\hangindent=2em\hangafter=1 #1.}

\newcommand{\referenceA}[3]{\hangindent=2em\hangafter=1 #1. #2, #3.}

\newcommand{\referenceB}[4]{\hangindent=2em\hangafter=1 #1. {\it #2}, #3, #4.}

\begin{document}
\title{Further understanding the interaction between dark energy and dark matter: current status and future directions}

\author[1]{B Wang \thanks{wang\_b@sjtu.edu.cn}}
\author[2]{E Abdalla \thanks{eabdalla@usp.br}}
\author[3]{F Atrio-Barandela \thanks{atrio@usal.es}}
\author[4]{D Pavón \thanks{diego.pavon@uab.es}}

\affil[1]{Center for Gravitation and Cosmology, Yangzhou University, China}
\affil[2]{Instituto de Fisica, Universidade de Sâo Paulo, Brazil}
\affil[3]{Física Te\'orica, Universidad de Salamanca, Spain}
\affil[4]{Departamento de Física, Universidad Autónoma de Barcelona, Spain}

%\begin{document}

\maketitle
%\title{Further understanding the interaction between dark energy and dark matter: current status and future directions}

%\maketitle
%\numberwithin{equation}{section}
%\numberwithin{table}{section}
%\numberwithin{figure}{section}

%\newpage
\def\reference#1#2#3#4#5{{\rm #1\;}{\rm #5\;}{\it #2\;}{\bf #3\;}{\rm #4}}
\def\referenceA#1#2#3{{\rm #1\;}{\rm #3\;}{\it #2}}
\def\referenceB#1#2#3#4{{\rm #1\;}{\it #2\;}{\rm #3}{\rm #4}}

\tableofcontents
%\newpage
%\section{introduction}
\begin{abstract}

The interaction between dark matter and dark energy can be incorporated into field theory models of dark energy that have proved successful in alleviating the coincidence problem. We review recent advances in this field, including new models and constraints from different astronomical data sets. We show that interactions are allowed by observations and can reduce the current tensions among  different measurements of cosmological parameters. We extend our discussion to include constraints from non-linear effects and results from cosmological simulations. Finally, we discuss forthcoming multi-messenger data from current and future observational facilities that will help to improve our understanding of the interactions within the dark sector.
\end{abstract}


\section{Introduction}
The description of the Universe has always been a major and deep problem in the History of Human understanding, as well as in the framework of Modern Physics based on Einstein Gravity. The cosmological solution, e. g. the Friedmann-Lema{\^\i}tre-Robertson-Walker (FLRW) solution was the first attempt to describe the Universe as a whole as a consequence of a physically well established theory. Since 1933 Dark Matter was introduced as a new ingredient, for a long time the question of whether there exists a hidden Dark Sector was a burden for a high level understanding of the Universe, since no natural representative from Elementary Particle Standard Model has ever represented such a new Cosmological sector. As a matter of fact, this is, in most aspects, a still standing problem. As we know it today, in the so called concordance model, the Dark Sector is constituted by two parts, Dark Matter (DM) and Dark Energy (DE), the former having the usual meaning of gravitationally interacting mass, but oblivious to other standard Particle Physics interactions, while the latter is a new strange kind of object leading to the accelerated expansion of the Universe.

On the other hand, the Standard Model of Particle Physics as described by Quantum Field Theory (QFT) is the most successful description of Nature. The Quantum Field Theory description of Quantum Electrodynamics is, experimentally, a theory with perfect accuracy. No scientific description of Nature can rival or substitute QFT. We take for granted that this is the case for Dark Matter. In the case of Dark Energy there is an alternative description, namely, a cosmological constant. Although much simpler, and thus favored in the sense of the Occam razor, there are difficulties concerning with the smallness of the Cosmological Constant or the fact that it effectively introduces ad hoc a new constant of Nature.

The existence of a DE component that is driving the acceleration of the Universe is well established \cite{Abdalla:2020,Abdalla:2022a,Bahamonde:2018,Huterer:2018}.
The simplest model to account
for the observed acceleration assumes DE is the cosmological
constant $\Lambda$.
While the $\Lambda$CDM concordance model accounts with great accuracy for cosmological observations 
over many length scales spanning more than 10Gyrs,
there are some important disagreements 
\cite{Planck_2018_1:2020}. If $\Omega_m$ denotes the current matter density
in units of the critical density and $\sigma_m$ is the mass rms fluctuation
on a sphere of $8h^{-1}$Mpc, with $h$ the Hubble constant in units
of $100$kms$^{-1}$Mpc$^{-1}$, the amplitude of 
galaxy clustering derived from weak lensing surveys 
and expressed by $S_8=\sigma_8\sqrt{\Omega_{m}/0.3}$ prefers lower values than 
the concordance model normalized to Planck at the $2-4\sigma$ confidence level
\cite{Heymans:2013, Hildebrandt:2017, Khlinger:2017, Planck:2016,Planck_2018_6:2020}.
The angular diameter distance at redshift $z\simeq 2$ derived from Ly-$\alpha$
measurements disagrees with the prediction of the concordance
model normalized to Planck at about $2.3\sigma$ \cite{Bautista:2017,
duMas:2017}. The amplitude of the matter power spectrum 
inferred from the abundance of rich clusters of galaxies also implies a lower value
although the discrepancy could be due to differences on the bias parameter derived from
simulations and inferred from observations \cite{Blanchard:2021}.

The most cited discrepancy between the measured cosmological parameters is the
high redshift estimate of the Hubble constant by Planck $H_0=(67.4\pm 0.5)$kms$^{-1}$
Mpc$^{-1}$ \cite{Planck_2018_6:2020}
and the traditional distance-ladder measurement by the
SHOES project $H_0=(74.03\pm 1.42)$kms$^{-1}$ Mpc$^{-1}$ \cite{Galvany:2022,Riess:2019}.
While the central values have remain stable in the past decade, the discrepancy
has grown from a $2.5\sigma$ in 2013 to $4.4\sigma$ in 2019. 
Systematic uncertainties related to the choice of Cepheid color-luminosity 
calibration could be behind the disagreement \cite{Mortsell:2021}
or the SNIa luminosities could evolve with redshift \cite{Ferramacho:2013}.
The constancy of the latter has been tested and current data is
inconclusive \cite{Benisty:2022}. This uncertainty leads to
degeneracies between the Hubble constant, DE parameters and
SNIa systematics \cite{DiValentino:2020c}.  It is necessary to reduce the
systematics of local measurements to clarify the Hubble
tension \cite{Dhawan:2022}. At the same time
it is worth noticing that the Hubble constant is not measured
from Planck data but it is a parameter derived using the concordance model. From the position
of the acoustic peaks in the power spectrum, the Cosmic microwave background (CMB) 
data determines the acoustic scale $\theta^*=r^*/D_A$, 
the ratio of the comoving size of the sound horizon at recombination $r^*$ to the 
angular diameter distance $D_A$. The
estimate is very robust and stable to the change of cosmological model.
In $\Lambda$CDM, the value of $\theta^*$  fixes $\Omega_mh^3$  
leaving a residual degeneracy between $h$ and $\Omega_m$ \cite{Planck:2014}.
Changing  $\Omega_mh^2$ 
changes the sound horizon and $D_A$ has to vary by a similar amount.
At low redshifts, $D_A\propto \int dz[\Omega_mh^2[(1+z)^3-1]+h^2]^{-1/2}$, 
and to keep $\theta^*$ constant $h$ must vary as well. This opens the possibility 
that models where the matter density evolves differently with redshift, 
$\Omega_m(z)\ne \Omega_m(z+1)^3$,
could be constructed that would resolve the tension.

Data tensions have motivated the search of different models as well as the analysis of the internal 
consistencies in the data. The discrepancies are largely due to values derived from CMB, i.e. high redshift, 
data with direct measurements at low redshifts. However, the data themselves could be affected by unknown systematics. 
For instance in \cite{Bhatta:2018}, $\Lambda$CDM, CDM and WDM with varying DE equation of state and interacting 
models were compared with observations, the latter fitting the data slightly better than the other models 
although the authors noted that the data sets were inconsistent among themselves. 
In \cite{Poulin:2018}, 
it was found that the tension could be
alleviated if Baryon Acoustic Oscillations (BAOs) or Supernovae Type Ia (SNIa) data were omitted
from the analysis.

If these tensions are confirmed it would signal the inadequacy of the $\Lambda$CDM to explain simultaneously the high and low redshift Universe. In the light of the new data, many models based on different physical considerations have been analyzed and have been recently reviewed. In \cite{Bahamonde:2018} a detailed application of dynamical systems to cosmological models, including models with interactions, is presented. The authors only analyzed the background evolution and discarded those models that did not present or had a transient late-time acceleration epoch. They did not analyze the evolution of density perturbations or compared model predictions with data. This comparison has been carried out in \cite{Schoneberg:2022} where models were ranked according to their ability to reduce the discrepancy between the high and the low redshift values of the expansion rate. The models considered did not explain satisfactorily both the $H_0$ and $S_8$ tensions. Also models with DM and DE interactions were not included. In \cite{DiValentino:2021b} the difficulties to fit the highly precise and consistent data at high and low redshifts were reviewed. Models with dynamical energy, interactions or modified gravity looked the most promising although no specific model looked far better than the others.  A summary of the difficulties of reconciling the standard $\Lambda$CDM cosmological with the
data is given in \cite{Perivolaropoulos:2022}. A thorough discussion of the current problems in Cosmology and future research directions is given in \cite{Abdalla:2022a}.

In this article we review the latest developments in the field of interacting dark energy (IDE) models, with a depth and cohesion not covered by the previous references. The present work is an update of \cite{Wang:2016} to which we refer the reader. Since our previous review, a large number of new interacting models have been constructed motivated by the shortcomings of the $\Lambda$CDM model. Here, we summarize phenomenological interacting DE models, and pay particular attention to interacting DE models based on a theoretical field description. These models have been less studied due to the intrinsic difficulty of deriving interaction kernels from first principles, but they are important since these models are based on a  field theory approach on the interaction between dark sectors.  Considering that current main tests of interacting DE models were limited to using the background dynamics and linear perturbation level related observational data, we generalize the discussion to the non-linear perturbations and examine the effect of DE and DM interaction on cosmic structure formation. We give a detailed summary of recent progress in the study of non-linear evolution and numerical simulations of structure formation in IDE models.  Most of the constraints on interacting DE models have come from large-scale and high-redshift observations, including cosmic expansion history, CMB, and large scale structures etc.  We require complementary probes to explore new physics. In here we discuss prospects to be available from future observations. We emphasize  new multi-messengers probes, including observations of the 21cm neutral hydrogen line and gravitational waves. These forthcoming observations will continue to give motivation to determine the nature of the dark sector. 

The present lack of data is such that motivations for an interaction beside theoretical adequacy are scarce. With more data about the variation of the equation of state with redshift can lead to an internal structure of the Dark Sector. 

Briefly, in Sec. 2 we summarize the most recent new phenomenological models, the new interacting kernels that have been introduced, their different motivations, their solutions to the $H_0$ and $S_8$ tensions and the forthcoming data sets that would help to refine the models. In \cite{1908} and \cite{2103}, the authors recently reviewed many proposals to resolve the Hubble tension, including some phenomenological interactions between DE and DM. In addition to phenomenological interacting DE models, in Sec. 3 we summarize interacting DE models based on theoretical field descriptions. These attempts are important to disclose deep physics of dark sectors and intrinsic relations between them. To examine interactions between DM and DE in light of the latest cosmological observations, especially their influence on the cosmic structure formation, in Sec. 4 we review recent advances in the study of non-linear evolution and numerical simulations of structure formation in IDE models. We skip the linear perturbation theory, which has been extensively discussed in \cite{Wang:2016}. In Sec. 5 we discuss the prospects to constrain model parameters using new observations and the forecast from future experiments. Finally, in Sec. 6 we present our conclusions.

\section{Recent progress in interacting dark energy models}

Models with dark matter interacting with dark energy were initially motivated
to solve or alleviate the coincidence problem, i.e., why the energy densities of 
DM and DE are of the same order precisely today 
(see \cite{Wang:2016} and references therein). 
In interacting models, the sum of dark matter $\rho_c$
and dark energy $\rho_d$ energy densities
is conserved but one component can decay into the other
\begin{equation}
\dot{\rho}_c+3{\cal H}\rho_c=Q,\qquad
\dot{\rho}_d+3{\cal H}(1+\omega_d)\rho_d=-Q,
\label{eq:background1}
\end{equation}
where $w_d=p_d/\rho_d$ is the equation of state for the DE and $Q$ the 
interaction kernel. Due to the unknown nature of dark matter
and dark energy, the kernel is usually assumed and a large variety of models
have been proposed.  Among the kernels considered we have \cite{An:2017,Beltran:2016,
Caprini:2016,Costa:2017,Feng:2016,Marcondes:2016,Sharov:2017,Yang:2016}
\begin{equation}
\begin{array}{lll}
Q_1={\cal H}\xi\rho_c,  &
Q_2={\cal H}\xi\rho_{d}, &
Q_3={\cal H}\xi(\rho_c+\rho_{d}), \\
Q_4={\cal H}\xi(\rho_c+(1+\omega_d)\rho_{d}), &
Q_5={\cal H}\xi\sqrt{\rho_{d}\rho_c}, &
Q_6={\cal H}\xi\frac{\rho_c\rho_{d}}{\rho_c+\rho_{d}},\\
\end{array}
\label{eq:kernels}
\end{equation}
where ${\cal H}=\dot{a}/a$ and the derivatives are taken with
respect to the conformal time. These kernels correspond to the
categories {\it I -- VI} and are given in Table~\ref{table1}.
The dimensionless parameter $\xi$ determines the strength of the interaction and can 
be positive or negative. For the kernel $Q_1$ of eq.~(\ref{eq:background1}) when $\xi>0$ there
is more DM in the past than at present, while for $Q_2$ there is less.
Depending on the interaction unphysical solutions
could appear, as $\rho_c$ could become negative in the distant future
\cite{Izquierdo:2018}. Furthermore, because dark matter has entropy 
($S_{c} \propto N_{c}$, where $N_{c}$ is the number of DM particles) and DE 
is likely $\Lambda$ or a scalar field in a pure vacuum state without entropy, 
the second law of thermodynamics could be violated in that case and the coincidence problem would worsen \cite{Chimento:2003}.

\begin{figure}[thbp]
    \centering
    \includegraphics[width=0.8\linewidth]{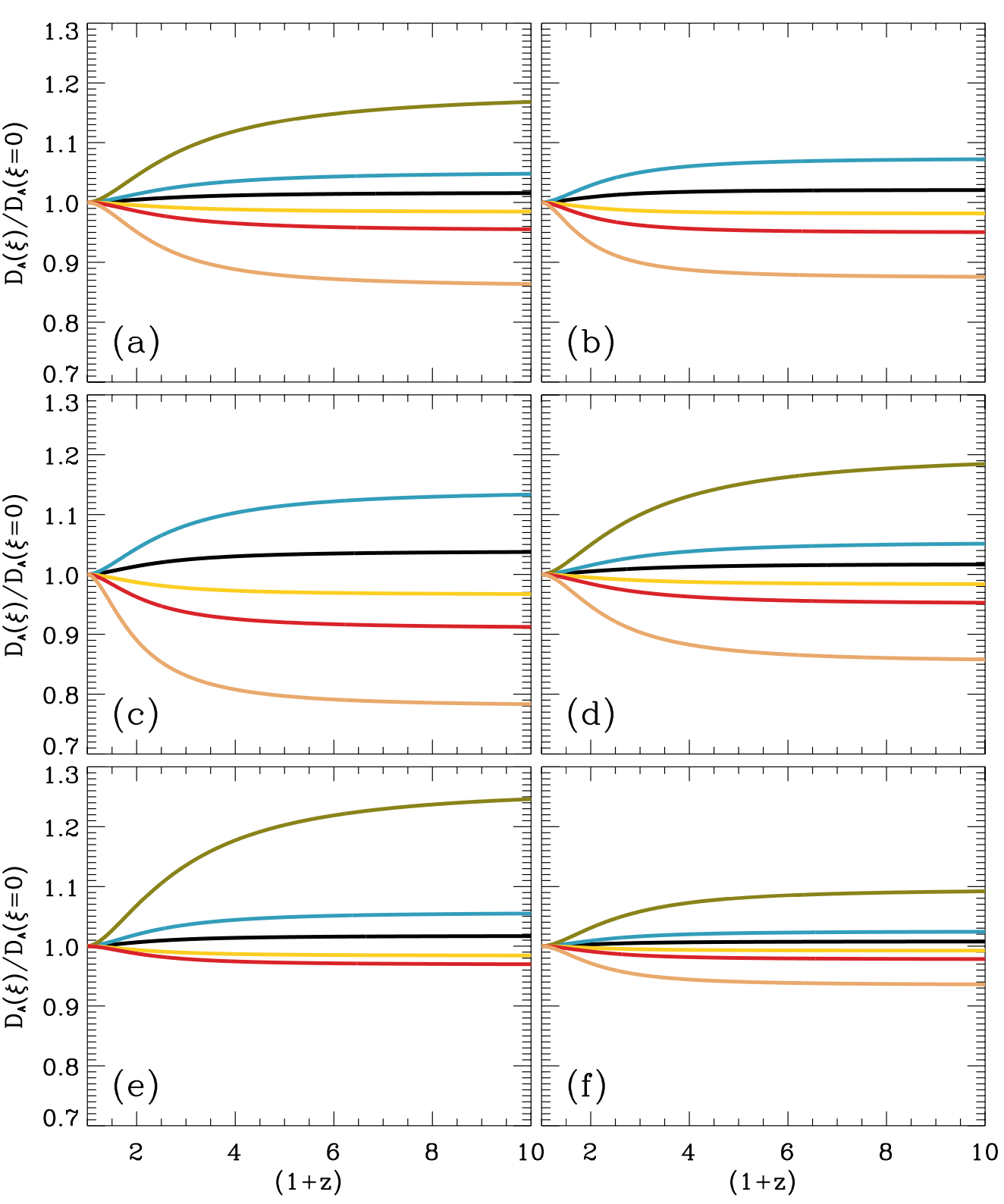}
    \caption{Angular diameter distance ratios for the six kernels given in eq. \eqref{eq:kernels} for a cosmological model with $\Omega_m=0.3, \Omega_{DE}=0.7$ and $\omega_x=-0.95$. Different colors correspond to different interaction strengths. In (a-d) and (f) black, turquoise, green, gold, red and brown correspond to interaction strength parameter $\xi=(0.1,0.3,1,-0.1,-0.3,-1)$, respectively. In (e) red corresponds to $\xi=-0.2$. Missing lines of any given color indicate the model was unphysical for that particular 
    value of the interaction parameter.}
    \label{fig:1}
\end{figure}

Although the initial motivation for IDE models was to solve or alleviate the coincidence problem,
more recently, the emphasis on these models has moved 
towards explaining the discrepancy between the measured value
of the Hubble constant derived from CMB data and the local measurements.
As the tensions on the value of the Hubble constant at high and low
redshifts remained as the data improved, IDE models
have been studied as potential candidates to explain away
the discrepancies between the $\Lambda$CDM model and observations.
As explained in the introduction, CMB data does not determine the value of the Hubble constant directly; 
it determines the acoustic scale $\theta^{*}$. In models where the fraction of DE is negligible at early times, 
the sound horizon at recombination $r^{*}$ would be the same and by varying the angular diameter distance $D_A$ 
one could expect to accommodate the measured values of $H_0$ at high and low redshift. 
More simply, CMB data constrains $\Omega_ch^2$ and since IDE models vary the matter density at early 
times, $h$ has to vary as well to fit the CMB constrain. In Fig \ref{fig:1} we 
represent the ratio of the angular diameter distance $D_A$ of the interacting model to the same model with no 
interaction. The six models represented correspond to the kernels given in eq \eqref{eq:kernels}. Colors represent 
different interaction strengths and missing lines correspond to unphysical solutions
that were not plotted. In these models the DE density at high redshifts is small and the angular diameter 
distance ratio remains constant at $z\gtrsim 10$. The interaction can change $D_A$ by as much as $20\%$, although to 
determine if any particular model alleviates the Hubble tension requires a detailed comparison with the data as 
summarized below.

A DM-DE interaction can also alleviate the $S_8$ tension
since it alters the growth of structure and, therefore,
changes the measure of gravitational clustering $S_8$. Matter
perturbation growth is determined by the free-fall time 
$t_{ff}\propto\sqrt{G\rho_c}$. DM density perturbations evolve
with an effective gravitational constant $G_{eff}$ that differs
from Newton's gravitational constant $G$ that determines the
evolution of baryons \cite{Olivares:2006}. $G_{eff}$ could be larger or smaller
than $G$ depending on the interaction kernel (see Sec~4.1). 
If $G_{eff}>G$ DM will cluster faster than baryons (galaxies) reducing the
discrepancy between the value derived from the CMB and from
lensing surveys. A detailed analysis would require not only the study of
the evolution of density perturbations in the linear regime but
also will need to follow the non-linear evolution with numerical
simulations . Notice that while a particular model can alleviate the
Hubble constant tension, it does not necessarily mean that it will solve the $S_8$ tension as well,
and vice versa (see below).

The dynamics at the background level, its critical
points and past and future singularities have been studied 
\cite{Deogharia:2021,Khyllep:2021,Pan:2020c,Panotopoulos:2020}; in some models
dark-matter observers would see no singularities \cite{Alvarez-Ortega:2022}.
In \cite{Biswas:2021}
the analysis of an interacting Umami Chaplygin gas showed the model alleviates
the coincidence problem. A review of analytic solutions of the Einstein-Friedmann
equations including scalar fields and interacting fluids is given in \cite{Faraoni:2021}.

Some theoretical effort has gone into providing a theoretical footing to
the interaction kernels, well using a scalar field description \cite{Cardenas:2020,Pan:2020b}
or deriving them from the continuity and Boltzmann equations \cite{Ludwick:2018,Ludwick:2021}.
In \cite{Johnson:2021a} it was demonstrated that a classical field and
a fluid description are consistent only for a very specific interaction kernel. For 
this model the data prefer a negative interaction parameter \cite{Johnson:2021b}.
Lacking a clear theoretical understanding, studies in this field differ in both
the considered interaction kernel and the observational data used to constrain model parameters.
The different kernels in eq.~(\ref{eq:kernels}) have been analysed and fit
to the data. Not all the kernels generate a valid cosmological
evolution \cite{Bernardi:2017} and the presence of an  interaction in
the dark sector modifies constraints on cosmological parameters like, for example,
the upper limits on neutrino masses \cite{Feng:2020a,Guo:2017,Yang79,Zhao:2020}.
In general, changing the type of interaction gives rise to a different cosmological
evolution at both the background and perturbation levels.

From the statistical point of view, interacting models enlarge the parameter space.
Information criteria are needed to establish when the improvement on fitting the data
is statistically meaningful \cite{Arevalo:2017}. Commonly used
criteria are the Akaike \cite{Akaike:1974}, Bayesian  \cite{Schwarz:1978} and
Deviance \cite{Spiegelhalter:2002} Information Criteria. Current results show that
models with additional parameters compared with the concordance $\Lambda$CDM model
are penalized by the conservative Bayesian Information Criteria \cite{Zheng:2017}.
%the following text was moved to a later point.
%Cosmic shear measurements from KiDS and Planck CMB data showed that interaction with
%kernels $Q_1$, $Q_2$, $Q_3$ alleviate the discordance with $\Lambda$CDM \cite{An:2018}.

\subsection{New phenomenological models}

Phenomenological models are usually constructed by introducing an energy
exchange between the dark components in the continuity equations
as given in eq~(\ref{eq:background1}). The interaction may couple
a generalized Chapligyn gas with DM \cite{Aljaf:2021}, DE with
neutrinos \cite{Amiri:2021}, different families of axions
\cite{Mawas:2021} or two scalar fields \cite{Sa:2021}.
The interaction could be effective at early or at late times
\cite{Sinha:2021}. IDE models and models with DM or DE self-interaction
have also been analyzed \cite{Balakin:2021}. The interaction can be
seen as a open thermodynamical system \cite{Harko:2022}. 
In \cite{Shafieloo:2018} the decay rate is constant, similar to radiactive decay.
Recently, models where DM and DE only exchange momentum have also been considered.
The coupling between DM and DE depends on their relative motion
and the interaction only modifies Euler equation.
By keeping  the continuity equation unaltered the background 
 evolution is identical to the concordance model
\cite{Beltran:2021a,Kumar:2017a,Linton:2022,Pourtsidou:2016,Simpson:2010}.
Couplings of DE to baryons \cite{Beltran:2020,Vagnozzi:2020}, 
to radiation \cite{deMartino:2018} and
to both DM and ordinary matter \cite{Harko:2022} have also been considered.

The underlying theory of gravitation can also be changed. Interacting models
in alternative theories like trace-free Einstein gravity \cite{deCesare:2022}
or gravity with torsion \cite{Izaurieta:2020} have been constructed. In the
former, the energy-momentum
tensor is not conserved and it can accommodate transfer of energy-momentum within
the dark sector with an equation of state $\omega_{DE}=-1$ for the DE. In the latter,
the torsion degrees of freedom in the dark sector are equivalent to an interaction.
In \cite{YangT:2020} the interacting model was based on the string Swampland criteria,
\cite{Akarsu:2020,Chagoya:2023} in the Rastall gravity extension of General Relativity,
while in \cite{Otalora:2022} the IDE arises from the matter spin coupling to gravity.

New models were defined that require the interaction to change sign by either introducing
a variable interaction kernel \cite{Abdollahi:2017,Arevalo:2019,Guo:2018a,Pan:2020a} or
a dependence with the first order derivatives of the energy densities
\cite{Pan:2018,Yang:2018c}. The evolution of the background and its future singularities
of the latter models have been analyzed by \cite{Aljaf:2020}. In \cite{Yang:2018a}
the kernel included a factor $(1+\omega_x)$. It was found that
the data preferred non-zero interaction although the no-interaction model
was within the 68.3\% confidence level. With similar kernels, \cite{Yang:2018b}
concluded that the $H_0$ tension was partially alleviated. In \cite{vonMarttens:2019}
study a more generic kernel than generalize those of eq.~(\ref{eq:kernels}).
These authors concluded that the range of the admissible parameters
was narrow, the non-interaction model being always consistent at the $1\sigma$ C.L.
In \cite{Paliathanasis:2019}, the interaction rate is a non-linear (quadratic) function
of the energy densities while in \cite{Yang:2019a} the interaction kernel contains
an exponential function of the DE to the DM ratio.
A Bayesian model selection analysis of linear and non-linear kernels
showed that only one third of the models considered fit the data as well
as the concordance model with the rest performing worse \cite{Cid:2019}.

The cosmological principle has also been applied to interacting models
and proposals have been made based on holographic considerations
\cite{Abdollahi:2019,Bouhmadi:2018,Cruz:2018,Cruz:2019,Cruz:2020,Escamilla-Rivera:2021,Feng:2018,Lepe:2016,Mukherjee:2016,Nayak:2020,
Sadri:2019a,Sharma:2020,Sinha:2020,Wang:2017}, providing some theoretical advantages.
%Other models combine the holographic principle with thermodynamic aspects
In \cite{Zarandi:2021} holographic models are constrained using the age of astrophysical
objects. For instance, for suitable choices of the interaction kernel, singularities that
are present in the holographic Ricci DE model are removed \cite{Belkacemi:2020}. However,
variants of the holographic model are strongly disfavoured by the data
\cite{Landim:2022,Sadri:2019b}.
Other degrees of freedom have been explored. Models have been constructed with the
interaction coupling, neutrino masses and relativistic degrees of freedom as free parameters
\cite{Kumar:2017b,Yang:2020a}.

In other works the interaction is not introduced through a kernel but it is described
as a modification of the time evolution of cosmological parameters such as
the DM density \cite{Aich:2022,Kumar:2016,Nunes:2016,Yang:2017} or the cosmic deceleration
\cite{Mukherjee:2016,Pradhan:2020} with respect to the standard
non-interacting $\Lambda$CDM model. Another approach is
to parametrize the modifications of the dark matter perturbations equations
in a model-independent way so different models can be described with the same
formalism and discriminate modifications of gravity from other effects
\cite{Aparicio-Resco:2021}. In \cite{Mukherjee:2021,Mukherjee:2022} the interaction is
not postulated but reconstructed from the data in a non-parametric way.

\subsection{Recent advances}

New data has helped to update the constraints on model parameters and
test for solutions of the $H_0$ and $S_8$ tensions. The most commonly
used data sets are Planck 2018 data release,
measurements of the late-time the expansion of the Universe,
Supernova Type Ia (SNIa), Baryon Acoustic Oscillations (BAOs),
cosmic shear and gravitational lensing, galaxy clustering 
and local distance ladder measurement of the Hubble constant.
Let us now briefly summarize some of the latest analyses.
Models based on kernels of eq.~(\ref{eq:kernels}) have been extensively
studied and compared with different data sets. In \cite{Pan:2017} the kernel $Q_3$ with constant and variable $\omega_d$ was examined and it was shown that IDE models provide similar fits to the data than $\Lambda$CDM, the latter being favoured
over interacting models by statistical information criteria due to smaller
number of free parameters.
Planck and Hubble constant data alone indicates the existence of a $Q_2$ coupling
at the 5.7$\sigma$ C.L. \cite{DiValentino:2021d}.
The limits of active neutrino masses and sterile
neutrino parameters depend on the DE properties \cite{Feng:2020b}.
In \cite{Sadri:2020,Yang:2019c} holographic models with interaction have been constrained
by the data and, while non-interacting scenarios are consistent with the
data, the interaction models are not ruled out.

The detection by the Experiment to Detect the Global Epoch of Reionization Signature
(EDGES) Collaboration of an excess in the 21-cm brightness temperature
compared with the expected value in $\Lambda$CDM \cite{Bowman:2018}
has been explained as a result of DM-DE interactions.
In \cite{Costa:2018,Mukhopadhyay:2021} the interaction model
decreases the Hubble parameter in the past and increases the optical depth
boosting the amplitude of the 21cm line. Similar results were found in
\cite{Xiao2019} although their model was in tension with other probes.
In \cite{LiC:2020} it was argued that reasonable changes in the optical depth and
Compton heating were easier to realize by introducing an early epoch
dominated by the DE. In \cite{Li:2019} the dark sector transfers heat to baryons to
explain the temperature excess and \cite{Halder:2021} analyzed the effect
of an interaction in the presence of primordial black holes.
The effect of DM-DE interactions
on the reionization process was further studied in \cite{An:2019}.

N-body codes have been constructed to study the growth and non-linear evolution of
the large scale structure for interacting models \cite{Zhang:2018} and use
the redshift evolution of the halo mass--concentration relation to
constrain the interaction parameter \cite{Zhao:2022}.
The results of these simulations were later compared with weak lensing
measurements. Since lensing is sensitive
to the non-linear evolution of scales, it could offer stronger constraints than
probes of the background expansion and linear evolution of density perturbations
\cite{Zhang:2019}. Other authors used an effective field theory
of Large Scale Structure to compute analytically the mildly non-linear evolution
to test the predictions of IDE models in this regime \cite{Nunes:2022}.
Among the non-linear probes that have been found to depend on the
DM-DE interaction are the formation of DM halos \cite{Liu2022}
or the ellipticity of cosmic voids \cite{Rezai:2020}.

\subsection{The tension in the $\Lambda$CDM model}

The multi-parameter cosmological models constructed recently aim to solve or alleviate the $H_0$ and $S_8$ tensions. Fitting IDE models to Planck, BAO and SNIa data has shown to reduced both $H_0$ and $S_8$ tensions \cite{DiValentino:2017,DiValentino:2020a,Kumar:2021}. In \cite{Gariazzo:2022} a prior on the absolute magnitude of SNIa was used instead of a prior on the Hubble constant and it was found that IDE models alleviate the $S_8$ tension between CMB and weak lensing data. In \cite{Lucca:2020,Zhai:2023} the Hubble tension was reduced to a $2.5\sigma$ C.L. and the statistical model comparison showed no preference of the concordance model over the interacting model. For a different kernel,  \cite{Yang:2018c} found that the evidence criteria still favored the standard model. Similarly, \cite{Yang:2022} found that energy exchange within the dark sector could alleviate both $H_0$ and $S_8$ tensions. The model required the strength of the interaction $\xi$ to vary with time. As a matter of fact, a coupling constant varying with time indicates a more sofisticated interaction, typical of a field theory model. In \cite{Lucca:2021} it was found that an interaction described by the $Q_2$ interaction kernel alleviated the $S_8$ tension without a significant impact on other cosmological parameters or statistical measures.

Different data sets lead to different conclusions. Data on the transversal BAO scale and Planck data predicts a Hubble constant in agreement with local observations and favors IDE at the $3\sigma$ CL \cite{Bernui:2023}. Time delay data from gravitational lenses in combination with other data sets showed that no IDE model could simultaneously alleviate the Hubble and $S_8$ tensions and improve the information criteria with respect to $\Lambda$CDM \cite{Wang:2022}. In models where only the Euler equation was modified, the data favored the interaction at the $3\sigma$ C.L., mostly driven by Planck cluster counts \cite{Figueruelo:2021}. These models have shown to alleviate the $S_8$ tension \cite{Beltran:2021b,Cardona:2022}
and predict a shift in the turn-around of the matter power spectrum and a bias between the baryon and DM peculiar velocities due to the interaction, opening new windows for detecting this coupling \cite{Asghari:2019}. Non-linear kernels have shown to be consistent with current observations, providing a better fit to redshift space distortion (RSD) data \cite{Cheng:2020}.

In \cite{Kumar:2019} Planck temperature and polarization data, KiDs lensing data (see Sec~\ref{Sec:2.4}) and local measurements of the Hubble constant showed a slight preference of IDE models over the concordance model. In \cite{Pan:2019a} it was shown that interacting models allow higher values of Hubble constant in agreement with local measurements although the concordance model was within the $2\sigma$ C.L. Models were the interaction reverses sign, IDE are preferred to the concordance model, the Hubble tension is alleviated but the $S_8$ tension still persists. In \cite{Harko:2022} the model gives an acceptable value of the Hubble parameter. In \cite{Yang:2020a} Planck 2018 data and local measurements of the Hubble constant indicated the coupling was non-zero at the $3\sigma$ C.L. and the value of $H_0$ was found to be compatible with local measurements.

The latest Planck observations \cite{Planck_2018_1:2020} preferred a positive curvature of the Universe at the 99\% confidence level \cite{DiValentino:2021a} driving the Hubble constant tension to the $5\sigma$ level, but including a DM-DE interaction reduced the tension to $3.6\sigma$ \cite{DiValentino:2021b}. Planck and SNIa data favor a closed Universe and DM-DE interaction at the 99\% C.L. \cite{DiValentino:2021c,Yang:2021a}. When compared with Planck 2018, BAOs and local measurements of $H_0$, non-linear couplings alleviate the $H_0$ tension and show evidence of interaction at the $2\sigma$ C.L \cite{Pan:2020a}.

Incorporating additional parameters such as neutrino masses, massive sterile
neutrinos and relativistic degrees of freedom to IDE models
alleviates the $H_0$ tension but does not solve it \cite{Guo:2021,LiH-L:2020}.
The parameters that characterize neutrino properties did
not correlate with the coupling parameter \cite{Yang:2020a}.
In \cite{Kumar:2017b} data on cluster abundances favored the existence of an interaction.  
Cosmic shear measurements from KiDs and Planck CMB data showed that interaction with either
kernels $Q_1$, $Q_2$, $Q_3$ alleviates the discordance with $\Lambda$CDM \cite{An:2018}.
However, the simultaneous
solution of both tensions have remained so far elusive \cite{Anchordoqui:2021}.
The new models must produce a shift in the central values of $S_8$ and $H_0$
instead of larger uncertainties that would reduce the tension
among data sets thanks to larger error bars to avoid being penalized by
model selection criteria \cite{deSa:2022}.

\begin{table}
\centering
\begin{tabular}{|l|l|l|}
\hline
& Kernel $Q\propto$ & References \\
\hline
I & ${\cal H}\xi\rho_c$  &
        \cite{Aljaf:2020,Aljaf:2021,An:2017,An:2018,Bachega:2020,Banerjee:2023,
        Begue:2019,Caprini:2016,Cardenas:2020,Carrasco:2023}\\
        && \cite{deCesare:2022,Feng:2016,Feng:2018,Feng:2019,Grandon:2019,Guo:2017,
        Guo:2021,Yang79}\\
        &&\cite{Guo:2018a,Halder:2021,Khyllep:2022,
        Kumar:2016,LiC:2020,Liu2022,Mishra:2023,Mukhopadhyay:2021}\\
        && \cite{Rezai:2020,Rodriguez:2023,Santos:2017,Wang:2022,Xiao:2021,
        Yin:2023,Zheng:2017,Zhang:2018}\\
        && \cite{Zhang:2019,Zhao:2020,Zhao2022} \\
        II& ${\cal H}\xi\rho_d$ &
        \cite{Aljaf:2020,Aljaf:2021,An:2017,An:2018,Anchordoqui:2021,
        Bachega:2020,Begue:2019,Bernui:2023,Califano:2023b,Caprini:2016}\\
        && \cite{Carrasco:2023,Deogharia:2021,DiValentino:2021c,
        DiValentino:2020z,DiValentino:2020c,DiValentino:2021d,
        Feng:2016,Feng:2019}\\
        && \cite{Feng:2020,Gariazzo:2022,Ghodsi:2023,Grandon:2019,Guo:2017,Yang79,Guo:2018a,
        Halder:2021}\\
        && \cite{Izquierdo:2018,Joseph:2023,Kumar:2017b,Kumar:2019,
        Kumar:2021,LiC:2020,Liu2022,Lucca:2021}\\
        && \cite{Marcondes:2016,Mukhopadhyay:2021,Nayak:2020,Nunes:2022,Sadri:2019a,
        Santos:2017,Sinha:2020,Xiao:2021}\\
        && \cite{Wang:2022,vanderWesthuizen:2023,Yin:2023,Yang:2016,Yang:2020a,
        Yang:2020b,Yang:2021a,Yang:2023}\\
        && \cite{Zhai:2023,Zheng:2017,Zhang:2018,
        Zhang:2019,Zhao:2020,Zhao2022,Zhao:2022} \\
III & ${\cal H}\xi(\rho_c+\rho_d)$ &
        \cite{Abdollahi:2019,Aljaf:2020,Aljaf:2021,An:2017,An:2018,Arevalo:2017,
        Bouhmadi:2018,Bachega:2020,Cardenas:2020,Costa:2017}\\
        && \cite{Feng:2016,Guo:2018a,Halder:2021,LiC:2020,Liu2022,Mukhopadhyay:2021,Sharma:2020,
        Xiao:2021}\\
        &&\cite{Xu:2018,Yin:2023,Zhang:2018,Zhang:2019} \\
IV & ${\cal H}(\xi_1\rho_c+\xi_2\rho_d)$ &
        \cite{Aljaf:2021,An:2019,Carrasco:2023,Cid:2019,Costa:2018,
        Costa:2019,Lepe:2016,Pan:2017}\\
        &&\cite{Pan:2020c,Sharov:2017,Yang:2018c,Zhang:2018} \\
V & ${\cal H}\xi\sqrt{\rho_d\rho_c}$ &
        \cite{Aljaf:2021,Feng:2016,Ong:2023} \\
VI & ${\cal H}\xi(1+\omega_d)^\alpha\rho_c^\beta\rho_d^\gamma(\rho_c+\rho_d)^\delta$ &
        \cite{Aljaf:2021,Arevalo:2017,Cheng:2020,Feng:2016,Feng:2020,
        Li:2017,Pan:2019a,Rodriguez:2023}\\
        &&\cite{Yang:2018b,Yang:2019b,Yang:2020b,Yang:2023} \\
 & Others &
         \cite{Aguilar-Perez:2022,Aich:2023,Akarsu:2020,Aljaf:2021,Aparicio-Resco:2021,
        Arevalo:2019,Balakin:2018,Balakin:2021,Bonilla:2022,Carrasco:2023}\\
        &&\cite{Chagoya:2023,Cid:2019,Cruz:2018,Cruz:2019,Cruz:2020,
            Dam:2019,Escamilla-Rivera:2021,Escamilla:2023} \\
        && \cite{Grandon:2019,Guo:2021,Harko:2022,Khyllep:2022,
        Kumar:2017b,Lepe:2016,Li:2019,deMartino:2018}\\
        && \cite{Marttens:2019,Mukherjee:2016,
        Mukherjee:2021,Mukherjee:2022,Nunes:2016,Ong:2023,Otalora:2022,
        Pan:2018}\\
        &&\cite{Pan:2020a,Pan:2023,Rodriguez:2023,Salzano:2021,
        Santana:2023,Shafieloo:2018,Sinha:2021,Xiao:2020}\\
        && \cite{Yang:2019c} \\
\hline
 & Momentum exchange &
        \cite{Asghari:2019,Beltran:2020,Beltran:2021a,Beltran:2021b,Beltran:2022,
        Cardona:2022,Figueruelo:2021,Kumar:2017a,Palma:2023}\\
\hline
\end{tabular}
\caption{Summary of the most used kernels in phenomenological models.}
\label{table1}
\end{table}

In Table~\ref{table1} we list the functional forms
of the most commonly used kernels in the phenomenological approach and
the references that have considered them. More complicated
kernels that are proportional to, or combine, the simpler ones listed
are also included. Kernels containing derivatives of the energy
density since those can be reduced to those
in the table by means of Friedmann equations.
The item {\it Others} refers to those kernels that can neither be reduced to
nor contain the previous functions. It also includes models that DE interacts
with Dark Radiation or parametric reconstructions of the
interaction kernel using data. The item {\it Momentum exchange} refers to those
kernels where DM and DE only exchange momentum and the background
evolution is identical to the concordance model. Finally, in Table 2 we will summarize in the item 
{\it Field theory} to list out references that derive the interaction
from a lagrangian based on field theoretical considerations.
Other works where DM or DE interact with neutrinos or other matter
components are not included in this review.

%\subsection{Prospects for future study in this field}
%\subsubsection{The Coincidence Problem}

\subsection{New Observational Data}\label{Sec:2.4}

More accurate observations and extended data sets are helping and will help
to further constrain IDE models and test whether the tensions described
above are due to new physics or caused by unforeseen systematic errors.
Among the extended and more refined data currently available we have

\begin{enumerate}

\item{} Cosmic microwave background temperature anisotropies.
Thanks to a better modeling and correction of systematic effects
polarization data from the High Frequency Instrument can be used
to analyze cosmological models. Planck 2018 legacy data includes full
likelihoods for the temperature and polarization spectra
\cite{Planck_2018_1:2020,Planck_2018_5:2020,Planck_2018_6:2020} for this
purpose.
\item{} Supernovae Type Ia samples. Modern surveys are currently detecting SNIa
at ever increasing rate. Different surveys are optimized to detect supernovae
at different redshifts and constrains on cosmological parameters benefit from
using all the available data. The Pantheon+ survey includes 1550 confirmed
SNIa in the redshift range $0\le 2.3$  \cite{Scolnic:2022}, expanding the earlier Pantheon compilation of 1048 SNIa, which itself
extended the Joint Light-curve Analysis (JLA, \cite{Betoule:2014}) compilation
of 740 supernoave in the range $0.01\le z \le 1.30$.

\item{} Baryon acoustic oscillations. Angular diameter distances
derived from BAOs also cover a wide range of redshifts and trace the expansion
history of the Universe. The 6dF Galaxy Survey (6dFGS) \cite{Beutler:2011},
the Main Galaxy Sample of Data Release 7 of Sloan Digital Sky Survey (SDSS-MGS)
\cite{Ross:2015},
the LOWZ and CMASS samples from the latest Data Release 12 (DR12) of
the Baryon Oscillation Spectroscopic Survey (BOSS) \cite{Gil-Marin:2016}
have effective redshifts $z_{eff}=0.106, 0.15, 0.32, 0.57$, respectively.
In addition, the correlation of quasars with $Ly-\alpha$ forest absorption
from BOSS has provided a measurement of the BAO scale along and across
the line of sight at $z=2.36$ \cite{Font-Ribera:2014}.

\item{} Redshift space distortions. The growth rate of structure
is commonly quantified by the matter density r.m.s. fluctuation within
a sphere $8h^{-1}$Mpc radius $\sigma_8$ times the rate of growth $f$ of matter
density perturbations, $f\sigma_8$. An updated compilation of data that
expands the range $0\le z\le 2$ can be found in \cite{Skara:2020}.

\item{} Weak lensing.  Cosmic shear is the gravitational lensing effect
on the shapes of galaxies as their light travels through the intervening structures
on the way to the observer. It is sensitive to
to the geometry and the growth of cosmic structure in the Universe.
Available data set are the Canada-France-Hawaii Lensing Survey (CFHTLens, \cite{Asgari:2017}),
the Hyper Suprime-Cam Subaru Strategic Program
(HSC, \cite{Hikage:2019}) and
the Kilo-Degree Survey (KiDS, \cite{Hildebrandt:2017,Hildebrandt:2020}).
Updated measurements of the cosmic shear from the Dark Energy Survey
(DES) from $\sim 100$ million galaxies spanning more than 4000 sq-deg on the sky
\cite{Amon:2022,Secco:2022}. The consistency of these data sets is
discussed in \cite{Amon:2023}.

\item{} Cosmic Chronometers (CC). Old and passively evolving galaxies can be used
to measure the Hubble parameter at different redshifts. A table with 30
measurements of the Hubble parameters and their errors is given in \cite{Moresco:2016}.

\item{} Local value of the Hubble constant. The most recent measurement of
$H_0$ from the SHOES project, using an updated sample of SNe Ia in the near-infrared
with distances to nearby galaxies measured using the Cepheid period-luminosity relations
is given in \cite{Galvany:2022}.

\item{} Cluster abundances. The abundance of galaxy clusters is rather
unsensitive to the geometry and mainly probes
the growth of cosmic structure \cite{Ruiz:2015}.
Data from X-ray and Sunyaev-Zeldovich observations
and cluster number counts are summarized in \cite{Bernal:2016}.
Errors are dominated by the uncertainty on estimating the cluster mass from
the observables.

\end{enumerate}

%\include{acronyms}
%\section{New interacting dark energy models}

%\subsection{New phenomenological models}

%\subsection{New field theory descriptions}

\section{New field theory descriptions}

%It is interesting to understand the interaction between DE and DM in field theory. Some previous attempts in this direction were summarized in \cite{Wang:2016}. Considering that quantum field theory has been proven as the most reliable tool to describe the interaction in nature, it is worth looking for a realistic quantum field theory description of IDE models. 
%%%%%%%%%%%%%%%%%%%%%%%%%%%%%%%%%%%%%

The difficulty in having a Field Theory description of the Dark Sector is connected to the fact that in the cosmological case one needs a dynamical and nonperturbative effect, what in field theory leads to difficulties. Non renormalizability is just one of them, though we are not concerned with quantum fields at this stage. The degree of non-linearity is a big issue. For Dark Energy one needs a nontrivial vacuum, with non-zero energy, what is, generally, a highly non trivial and unstable possibility in QFT. Nonetheless, this is a path worth following, and it has been performed with some degree of understanding.

There are two main problems connected to the Dark Sector. We need a fundamental description for it, and the only well known description of fundamental matter up to now is field theory, what we intend to argue for below. In this framework the so called Dark Fields below belong to a set of new fundamental fields not in interaction with QED, and possibly not either to the Standard model fields, although a description of Dark Matter in terms of a chargeless field still belonging to the Standard Model is not excluded, and in fact is a part of the description of a Field Theory of the Dark Sector.

A new problem however arises in this framework, especially concerning models whose Dark Energy is still described by 
a Cosmological Constant, namely the smallness of its value, as analysed long ago by Weinberg \cite{Weinberg}. This will be briefly described in one of the next subsections.

\subsection{Dark Sector Field Theories}

Let us be concerned about the case of the Dark Sector as a whole. The obvious difficulty is that no experience exists about the structure of the Dark Sector besides its very existence. However, there is an actual starting point: if the Dark Matter is described by Quantum Field Theory, interaction is mandatory, especially in case of a Lagrangian description, since any allowed term in a Lagrangian is mandatory as well, including an interaction of the subsectors. Therefore, we arrive quite naturally at a new Sector, with no electric charge (hence Dark to our eyes) but with possible enormous degree of complexity and interactions. The case of Dark Energy still remains, namely, can it be described by a small Cosmological Constant?
We can summarize the question in terms of three/four main classes:

1.	Simple Field theoretic models for Dark Energy and Dark Matter in interaction.
2.	Dark Matter as a Standard Model Field, naturally invoking the Lightest Supersymmetric Particle. Dark Energy as a Cosmological Constant.
3.	Dark Energy and Dark Matter as part of a larger setup, possibly consequence of a larger Supergravity of Superstring theory. Or else, as a consequence of modified gravity.

In either the first or third cases above, interaction is part of the Dark Sector, while in the second case only Dark Matter may be understood in terms of a field theory description, but still, including an interaction with other fields except for Electromagnetism. The Cosmological Constant describes the acceleration of the Universe, a fact corroborated by observations. There are two basic theoretical problems in the Cosmological Standard Model. The first is the fact that $\Lambda$ is too small as compared to any Field Theory attempt to describe it in terms of Standard Field Theory. This is a question discussed in detail in the literature \cite{Weinberg}. The question is that it is very natural, in standard field theory to expect a very large cosmological constant, so large that a world as the one we live in cannot be conceived. Several possibilities have been argued and discussed, including supersymmetry and supergravity, but no definite answer exists so far. We are not able to decide in favor of any answer to this question, thus we basically ignore it and think that either the Cosmological Constant is indeed very small, by some magic, describing the Dark Sector or it vanishes by some symmetry and its effect is due to the inclusion of other modifications of the theory, such as new fields or a modification of gravity.

 The second basic question is that both DM and DE are equally important at the present era in spite on their different behavior as predicted by the Standard Cosmological Model. This is the coincidence problem. Therefore, it is natural to consider models of DM and DE with non-trivial characteristics, that is, not just a fluid and a constant. This is a fundamental question that should lead to a formulation of a more fundamental theory of these elements, possibly increasing the scope of the Standard Model of the Elementary Particles.
A common attempt to describe the Dark Sector independently of new fundamental physics is through the Horndeski models \cite{Zumalacarregui:2013pma}, \cite{Xiao2019}.

A more standard proposition relies of models describing DE/DM as canonical fields with mutual interaction. They rely basically on quintessence models of Dark Energy with a possible interaction with new matter fields. A prototype is a fermion-scalar model with a Yukawa coupling. In particular, this is compatible with recent discussions concerning string inspired models, what in \cite{Cicoli:2023opf} has been called Dark Matter assisted Dark Energy, or else, interacting Dark Sector. It is also claimed to be a scenario consistent with quantum gravity \cite{Cicoli:2023opf}. 

Some further interacting models based on Standard Field Theory are consistent with existing Lagrangians and minimal interactions, such as \cite{LandimBernardiAbdalla} and \cite{Landim:2016isc} 
Another intermediate step of attempting towards a field theory description via interacting quintessence is \cite{Boehmer:2015kta}.

Effects of DE and DM in interaction have been and are intensively in the course of being analysed via observations \cite{Abdallaetall2023}.
In the framework of string related models, disformal couplings or even non canonical Lagrangians have been considered. The tachyon field has been considered in \cite{Micheletti:2009pk}.

Some models are directly related to branes as in \cite{Koivistoetall2014}.
In general, these models are related to disformal transformations and directly lead to non conservation of the energy of Dark Matter, therefore interacting theories.

One of the common candidates to describe the Dark Sector is the axion, see \cite{Marsh:2015xka} for a review. Axions are certainly interacting particles and belong to the set of Standard Model fields. Hence this is a direction of the description of Dark Sector where the Standard Model itself is the source that generates the Dark Matter.

%%%%%%%%%%%%%%%%%%%%%%%%%%%%%%%%%%%%%%%%
The key obstacle to formulate a quantum field theory of DE interacting with DM comes from the decoupling, indicating that the heavy DM field cannot interact efficiently with light DE field. Naively introducing strong couplings between light DE field and heavy DM particles can induce extra long-range forces, which can  be ruled out easily.     

In \cite{Kaplor2016}, a model with multiple axions was introduced. Similar ideas were considered in inflation analysis, for example see \cite{Starobinsky1985, Starobinsky1992, Dimopoulos2008}. Some of the axions are assumed light enough describing DE. DM could be a combination of many separate components. Besides sufficiently heavy axions accounting for most decoupled DM, a fraction of DM can also be in the form of ultralight boson condensates of masses not too much greater than the current Hubble scale $H_0$. The mass of this part of DM is possible close to that of DE to evade decoupling. They can have significant interaction, in consistence with quantum field theory. 

The complete action of the multiple fields model is presented as \cite{Kaplor2016}  
\begin{equation}
S=S_g(g_{\mu\nu}) + S_{\phi}(\phi_1,\phi_2) + S_m, \nonumber
\end{equation}
where $S_g$ is the Einstein Hilbert action, $S_m$ is the action of the non-interacting CDM.  $S_\phi$ describes the interacting ultralight axion fields,
\begin{align*}
 S_{\phi}(\phi_1,\phi_2)\nonumber 
 =-\int d^4x\sqrt{-g}  \Big(\frac{1}{2}(\partial\phi_1)^2 + \frac{1}{2}(\partial\phi_2)^2  + V(\phi_1,\phi_2)\Big).
\end{align*}
$\phi_1$ and $\phi_2$ describe part of the DM and DE, respectively,containing comparable mass scale
\begin{equation}
 V\left(\phi_1,\phi_2\right)= V_1(\phi_1)+ V_2(\phi_2) 
+ V_{\text{int}}(\phi_1, \phi_2), \label{pot}
\end{equation}
where
\begin{align}
   V_1(\phi_1)=   \mu_1^4\left[1-\cos\left(\frac{\phi_1}{f_1}\right)\right], \,
   V_2(\phi_2)= \mu_2^4\left[1-\cos\left(\frac{\phi_2}{f_2}\right)\right], 
\end{align}
and the interaction term between $\phi_1$ and $\phi_2$ is given by 
\begin{equation}
V_{\text{int}}\left(\phi_1,\phi_2\right)=\mu_3^4\left[1-\cos\left(\frac{\phi_1}{f_1}-n\frac{\phi_2}{f_2}\right)\right]. \label{V_int}
\end{equation}

The total energy momentum tensor consisting of the non-interacting part of the CDM and the two fields $\phi_1,\,\phi_2$ from the variation of $S_m + S_{\phi}(\phi_1,\phi_2)$ with respect to the metric read
\begin{equation*}
T_{\mu\nu}=-\frac{2}{\sqrt{-g}}\frac{\delta(S_m+S_{\phi})}{\delta g^{\mu\nu}}\,.
\end{equation*}

The equations governing the evolution of the universe 
are as follows 
\begin{align}
& {3H^2} = \rho + \frac{1}{2}\dot{\phi_1}^2 + \frac{1}{2}\dot{\phi_2}^2 \nonumber \\
&  + 2\mu_1^4\sin^2\left(\frac{\phi_1}{2f_1}\right)   +2\mu_2^4\sin^2\left(\frac{\phi_2}{2f_2}\right) \nonumber \\
& +2\mu_3^4\sin^2\left(\frac{\phi_1}{2f_1}-n\frac{\phi_2}{2f_2}\right)\, , \label{Friedmann}
\end{align}
and the Raychaudhuri equation is given by 
\begin{equation}
\label{Raych}
2\dot{H}=-\rho-\dot{\phi_1}^2 - \dot{\phi_2}^2\,. 
\end{equation}
Klein-Gorden equations are derived as 
\begin{subequations}
\label{KG}
\begin{align}
\ddot{\phi_1}+3H\dot{\phi_1}+\frac{\mu_1^4}{f_1}\sin\left(\frac{\phi_1}{f_1}\right) +\frac{\mu_3^4}{f_1}\sin\left(\frac{\phi_1}{f_1}-n\frac{\phi_2}{f_2}\right)=0, \label{KG1}\\
\ddot{\phi_2}+3H\dot{\phi_2}+\frac{\mu_2^4}{f_2}\sin\left(\frac{\phi_2}{f_2}\right)   -n\frac{\mu_3^4}{f_2}\sin\left(\frac{\phi_1}{f_1}-n\frac{\phi_2}{f_2}\right)=0. \label{KG2}
\end{align}
\end{subequations}
The continuity equation of the decoupled component satisfies 
\begin{equation}\label{cont-eq}
\dot{\rho}+3H\rho=0\,.
\end{equation}

This model involves a natural and consistent construction of an effective field able to drive  cosmic acceleration with possible DE interacting with a DM component. The dynamical system of this model was studied carefully in \cite{Saikat2021}. It was shown that the time-dependent system has the same dynamics as its corresponding time-averaged version. More complete analyses, using different astronomical observational data to constrain cosmological parameters in such model both in linear and nonlinear studies are called for. 

%%%%%%%%%%%%%%%%%%%%%%%%%%%%%%%%%
%%%%%%%%%%%%%%%%%%%%%%%%%%%%%%%%%%

Modifications of gravity are also a source of discussions concerning the possible analytical description of the Dark Sector, especially $f( R)$ theories, \cite{DeFelice:2010aj}
This, however, has been source of questions, since some possible inconsistencies have been found for $f(R)$ theories \cite{Amendola:2006kh}
Further possibilities of extended theories of gravity are found in \cite{Capozziello:2011et}

Nonetheless, in some generality, some of these theories can be related to interacting theories \cite{He:2011qn}

The idea that either a Dark Sector or even an accepted value of the Cosmological Constant can arise from the concept of swampland, derived from ideas concerning String Theories. These effective theories are explained in detail in \cite{Palti:2019pca} 
A more direct description of a possible de Sitter world (the socalled de Sitter conjecture on the swampland) is also discussed elsewhere, see \cite{OoguriPaltiShiuVafa2019} and also \cite{BoussoPolchinski2000}

As a matter of fact, one of the best hopes towards a fundamental theory relies on the string idea. Although today far from getting any hint from observations, even in cosmology, the ideas behind can propel discussions and hopes, such as \cite{Nojiri:2006je}
Some models contemplate a more involved internal structure of the Dark Sector \cite{LandimBernardiAbdalla} 
In fact, this is an attempt for finding a more complete and sound description of a full Dark Sector, an enterprise also connected to a description based on String Theory, as e. g. the $E_8\otimes E_8$ version of superstring theory, which is naturally divided in two Sectors, mutually blind, except for gravity, a situation we look for in the case of the Dark Sector description. There are some discussions of such a model in the literature \cite{Abdalla:2020,Honecker:2016,Acharya:2018} as a means to describe the Dark Sector. Our aim here is to review such descriptions and describe possible and theoretically sound ways to have a Field Theoretical description of the Dark Sector.

In order to include a Dark Sector we are obliged to go beyond the Standard Model in order to avoid charge as well as unstable states: we need Dark Matter (no interaction with Electrodynamics) and stability, i.e., a stable Dark Sector.

Some of the most appealing models are those representing Supersymmetry of even Superstring Theory, where, in basically both cases, several new chargeless particles exist, and a stable lightest (namely stable) particle may be available. In this case, for Dark Energy one does not seek for a field theory description, just for Dark Matter. The advantages are to have a full Field Theory description and basically no need of going beyond the Standard Model, but just accommodate Dark Matter as one of the Supersymmetric partners unknown today.

Today, cosmology is possibly the most important area of research in fundamental physics. Looking at a laboratory at least 13 orders of magnitude larger than our local planetary surrounding, with new fields and particles fulfilling circa 20 times more energy content than what is presented to us on earth, cosmology can provide us a view of the new. In case the Dark Sector has the same richness as the Baryonic Sector, a real possibility in the case of string inspired models, DE and DM is a tantalizing information, certainly beyond our common sense knowledge. In particular, DE contradicts the Strong Energy Condition, a behaviour quite unthinkable until recently. In view of these properties, it is no doubt that the physical properties of a purported Dark Sector are very peculiar and encompasses non trivial modifications of Standard Particle Theory and models. Previous important results seem to pale before these properties.

To summarize, In Table~\ref{table2} we list recent IDE models derived from different
lagrangians.

\subsection{The Cosmological Constant Problem}

It is now well known and quite established that the cosmological constant constitutes one of the least understood problems in fundamental physics. Indeed, in the case where one naively compares the computed vacuum energy from Standard model and the observational value of the cosmological constant, we have a factor of up to $ 10^{120}$ difference. Despite the fact that some arguments can ameliorate such a discrepancy, it means that our understanding of the question is essentially null! 

There have been a few attempts to argue about the smallness of the cosmological constant. It has already been argued that scale invariant field theories including gravity there are quantum Scale Invariant theories with naturally vanishing cosmological constant with a Higgs field playing the role of the dilaton. The authors claim that the model can describe Dark Energy as well\cite{Shaposhnikov:2008xi}. Using string theory ideas for the solution of the Cosmological Constant problem is now a well known idea \cite{Rubakov:1983bz}. A different approach is the idea of dynamical equations \cite{Armendariz-Picon:2000nqq} and also dynamical Dark Energy \cite{Sahni:2002kh} and even non-local theories \cite{Arkani-Hamed:2002ukf}

The fact is that several attempts have been tried with no really definitive solution. What is sure is that most of the approaches lead to a kind of dynamical cosmological, namely really Dark Energy, besides, of course, anthropic approaches that have certainly been tried, especially in the framework of string approach.
\begin{center}
\begin{table}
\begin{tabular}{|l|l|}
\hline
Field theoretical models &
        \cite{Akarsu:2023,Akarsu:2018,Alvarez-Ortega:2022,Amiri:2021,An:2019b,
        Arcia:2023,vandeBruck:2018,vandeBruck:2020,Cardona:2023,Chamings:2020}\\
        & \cite{Chibana:2019,Dusoye:2021,Dusoye:2023,Dutta:2018,Gomez:2023,
        Johnson:2021a,Johnson:2021b,Karwan:2017}\\
        & \cite{Kase:2020,LandimBernardiAbdalla,Liu:2023,Liu:2023a,Lucca:2020,
        Mandal:2023,Mifsud:2017,Nakamura:2019}\\
        & \cite{Paliathanasis:2019,Pan:2019b,Pan:2020b,Patil:2023,
        Roy:2019,Roy:2023,Sa:2021,Tao:2020,Thipaksorn:2019}\\
\hline
\end{tabular}
\caption{List of IDE models based on field theoretical considerations.}
\label{table2}
\end{table}
\end{center}

%%%%%%%%%%%%%%%%%%%%%%%%%%%%%%%%%%%%%%
%%%%%%%%%%%%%%%%%%%%%%%%%%%%%%%%%
%\subsection{New dynamical analysis}
%\subsection{Holographic Models}
%\subsection{Lagrangian Based Models}
%\subsection{Field Theoretic Formulation of DM/DE Interaction}
%\subsection{Challenges for Scaling Cosmologies}

%\include{Chapter_3_Current_constraint}
%\include{linear_perturbations}
\section{Cosmological simulations, Nonlinear structure formations}
%\subsection{The necessity of cosmological simulations and nonlinear constraints}

%\subsection{Weak lensing}
%\subsection{Redshift distortions}

%\subsection{Phenomenological Model}

Cosmological N-body simulations have been widely applied to study the nonlinear evolution of the large scale structure in the Universe \cite{FW2012,AH2022}.  In contrast to the standard $\Lambda$CDM model, once interactions within the dark sector are turned on, in addition to modifications in density perturbations, there are additional forces exerted on the dark matter particles. Special treatments are needed to simulate the structure formation in IDE models. Some attempts to build simulation codes for these models were made in \cite{Baldi2010, Baldi2011, Baldi2012, Carlesi2014a, Carlesi2014b, Penzo2016, Huillier2017}. Recently, a fully self-consistent numerical simulation pipeline was developed in \cite{Zhang:2018}.  This new code was used to simulate structure formation down to redshift $z=0$ and constrain the interaction between dark sectors by comparing the simulations with the SDSS galaxy-galaxy weak lensing measurements \cite{Zhang:2019}. The simulations have also been used to systematically study halo properties in IDE models \cite{Liu2022}.  High resolution cosmological N-body simulations covering a large range of the parameter space were used recently 
in \cite{Zhao2022} to examine the halo concentration-mass relation.  

\subsection{N-body simulations for interacting DE models}

Let us review briefly the main characteristics of the fully self-consistent IDE simulations. Detailed descriptions can be found in \cite{Zhang:2018}.      
In IDE models,  the covariant description of the energy-momentum transfer between DE and DM is given by \cite{Wang:2016}
\begin{equation}
\label{eq.Tdmde}
\bigtriangledown_{\mu}T^{\mu\nu}_{(\lambda)}=Q^{\nu}_{(\lambda)},
\end{equation}
where $Q^{\nu}$ denotes the interaction between two dark components and $\lambda$ denotes either the DM or the DE sector. For the whole system, the energy momentum conservation still holds, satisfying
\begin{equation}
\label{eq.Ttotal}
\sum_{\lambda}\bigtriangledown_{\mu}T^{\mu\nu}_{(\lambda)}=0.
\end{equation}
We concentrate on the general stress-energy tensor 
\begin{equation}
\label{eq.T}
T^{\mu\nu}=\rho U^{\mu}U^{\nu}+p(g^{\mu\nu}+U^{\mu}U^{\nu}).
\end{equation}
The zero-component of Eq. (\ref{eq.Tdmde}), namely (\ref{eq:background1}) provides the background conservation equations for the energy densities of the dark sector.
%\begin{gather}
%\label{eq.rhodm}
%\rho'_{c}+3\mathcal{H}\rho_{c}=a^2Q_{(c)}^0=Q,\\
%\label{eq.rhode}
%\rho'_{d}+3\mathcal{H}(1+w_d)\rho_{d}=-a^2Q_{(d)}^0=-Q,
%\end{gather}
%where the subscript ``c" denotes DM and ``d" denotes DE. %$\mathcal{H}$ is the Hubble function defined as %$\mathcal{H}=a'/a$, $a$ is the cosmic scale factor and the %prime is the derivative with respect to the conformal time, 
The interaction kernel is phenomenologically expressed as a  linear combination of the energy densities of dark sectors in the form of $Q=3\xi_{1}\mathcal{H}\rho_{c}+3\xi_{2}\mathcal{H}\rho_{d}$, where $\xi_1$ and $\xi_2$ are free parameters to be determined from observations. 

The perturbed space-time is given by
\begin{align}
\label{eq.ds}
ds^2=&a^2(\tau)[-(1+2\psi)d\tau^2+2\partial_iBd\tau dx^i \notag \\
&+(1+2\phi)\delta_{ij}dx^idx^j+D_{ij}Edx^idx^j],
\end{align}
where $\psi$, $B$, $\phi$, and $E$ represent the scalar metric perturbations, and $D_{ij}=(\partial_i\partial_j-\frac{1}{3}\delta_{ij}\bigtriangledown^2)$. 

The linear perturbation equations of IDE models were derived in \cite{he2}. 
The gauge invariant gravitational potentials, density contrast, and peculiar velocity are described as follows,
\begin{align}
&\Psi=\psi-\frac{1}{k}\mathcal{H}(B+\frac{E'}{2k})-\frac{1}{k}(B'+\frac{E''}{2k}), \\
&\Phi=\phi+\frac{1}{6}E-\frac{1}{k}\mathcal{H}(B+\frac{E'}{2k}), \\
&D_{\lambda}=\delta_{\lambda}-\frac{\rho'_{\lambda}}{\rho_{\lambda\mathcal{H}}}(\phi+\frac{E}{6}),\\
&V_{\lambda}=v_{\lambda}-\frac{E'}{2k}.
\end{align}
Choosing the Longitudinal gauge by defining $E=0$, $B=0$, we have
\begin{align}
&\Psi=\psi, \\
&\Phi=\phi, \\
&D_{\lambda}=\delta_{\lambda}-\frac{\rho'_{\lambda}}{\rho_{\lambda}\mathcal{H}}\Phi,\\
&V_{\lambda}=v_{\lambda}.
\end{align}
Considering the phenomenological form of the energy transfer between dark sectors defined above, we obtain the general gauge invariant perturbation equations for DM and DE respectively,
\begin{subequations}
\begin{align}
\label{eq.DUdm}
D'_c=&-kU_c+6\mathcal{H}\Psi(\xi_1+\xi_2/r)-3(\xi_1+\xi_2/r)\Phi' \notag \\
&+3\mathcal{H}\xi_2(D_d-D_c)/r,  \\
U'_c=&-\mathcal{H}U_c+k\Psi-3\mathcal{H}(\xi_1+\xi_2/r)U_c, \notag
\end{align}
\begin{align}
\label{eq.DUde}
D'_d=&-3\mathcal(C_e^2-w_d)D_d-9\mathcal{H}^2(C_e^2-C_a^2)\frac{U_d}{k} \notag \\
&+[3w'_d-9\mathcal{H}(w_d-C_e^2)(\xi_1r+\xi_2+1+w_d)]\Phi \notag \\
&+3(\xi_1r+\xi_2)\Phi'-3\Psi\mathcal{H}(\xi_1r+\xi_2) \notag \\
&-9\mathcal{H}^2(C_e^2-C_a^2)(\xi_ir+\xi_2)\frac{U_d}{(1+w_d)k} \notag \\
&-kU_d+3\mathcal{H}\xi_1r(D_d-D_c), \\
U'_d=&-\mathcal{H}(1-3w_d)U_d+3(C_e^2-C_a^2)\mathcal{H}U_d \notag \\
&-3kC_e^2(\xi_1r+\xi_2+1+w_d)\Phi+kC_e^2D_d \notag \\
&+3\mathcal{H}(C_e^2-C_a^2)(\xi_1r+\xi_2)\frac{U_d}{1+w_d} \notag \\
&+(1+w_d)k\Psi+3\mathcal{H}(\xi_1r+\xi_2)U_d, \notag
\end{align}
\end{subequations}
where $U_{\lambda}=(1+w_{\lambda})V_{\lambda}$, $C_e^2$ is the effective sound speed of DE, $C_a^2$ is the adiabatic sound speed, and $r=\rho_c/\rho_d$ is the energy density ratio of DM and DE.

From the perturbed Einstein equations, we can obtain the Poisson equation in the subhorizon approximation \cite{he2}
\begin{equation}
\label{eq.Poisson}
-k^2\Psi=\frac{3}{2}\mathcal{H}^2[\Omega_c\bigtriangleup_c+(1-\Omega_c)\bigtriangleup_d],
\end{equation}
where $\bigtriangleup_{\lambda}=\delta_{\lambda}-\frac{\rho'_{\lambda}}{\rho_{\lambda}}\frac{V_{\lambda}}{k}$, $\Omega_{\lambda}=\frac{\rho_{\lambda}}{\rho_{crit}}$, and $\rho_{crit}$ is the critical density. This is the bridge between the matter perturbations and the metric perturbations. The Poisson equation can be rewritten in real space as
\begin{equation}
\label{eq.RPossion}
\bigtriangledown^2\Psi=-\frac{3}{2}\mathcal{H}^2[\Omega_c\bigtriangleup_c+(1-\Omega_c)\bigtriangleup_d].
\end{equation}
The second equation in (\ref{eq.DUdm}) can give the velocity perturbation for DM of the form
\begin{equation}
\label{eq.vdm}
V'_c+[\mathcal{H}+3\mathcal{H}(\xi_1+\xi_2/r)]V_c-k\Psi=0.
\end{equation}
Combining  (\ref{eq.RPossion}), in real space we have the modified Euler equation
\begin{align}
\label{eq.Euler}
&\bigtriangledown V'_c+[\mathcal{H}+3\mathcal{H}(\xi_1+\xi_2/r)]\bigtriangledown V_c \notag \\
&+\frac{3}{2}\mathcal{H}^2[\Omega_c\bigtriangleup_c+(1-\Omega_c)\bigtriangleup_d]=0.
\end{align}
The coupling between dark sectors modifies the gravitational potential and further exerts an additional acceleration onto the DM particles. 

Phenomenological interacting models  were  investigated in \cite{Costa:2017}  by employing observational data sets including CMB data from Planck,  SNIa, BAOs and Hubble constant measurements. These numerical fitting results can be used as input parameters for the simulations to study the formation of structure. 

%\subsection{ME-Gadget Code}
Comparing with the cosmological simulations in the standard $\Lambda$CDM model, there are four modifications required when considering IDE models \cite{Baldi2010,Baldi2011}. First, the universe expansion history has been modified so that the Hubble diagram $H(a)$ should be explicitly given. 
Second, because of the energy flow between DM and DE, the mass of the DM particles $m(a)$ in simulations should changed as a function of the scale factor. Third, from the modified Euler equation DM particles  
receive an additional acceleration proportional to its 
velocity $\mathbf{a_v}=\alpha (a)\mathbf{v}$, where 
\begin{equation}\label{eq:alphaa}
\alpha (a)=-3\mathcal{H}(\xi_1+\xi_2/r)a.    
\end{equation} 
There is an additional minus sign and a scale factor $a$ in contrast to Eq.(\ref{eq.Euler}). They arise from a coordinate transformation \cite{Baldi2010}. $\mathbf{a_v}$ is referred to as the dragging force or friction term, although it is not necessarily slowing down the particles. Fourth and final, the gravitational constant $G$ is also different from that in the $\Lambda$CDM model. The additional force exerted on DM particles is also called the fifth force.  Eq. (\ref{eq.Poisson}) implies that the fifth force is caused by the perturbation of DE, which results in the modification of the Poisson equation in harmonic space $-k^2\Psi=\frac{3}{2}\mathcal{H}^2 \Omega_c\bigtriangleup_c(1+\beta(a,k))$, where $\beta(a,k)=(1-\Omega_c)\bigtriangleup_d /\Omega_c\bigtriangleup_c$. In \cite{Baldi2010,Baldi2011}, $\beta(a,k)$ was simplified as a constant. This however, is not accurate 
enough to capture the distributions of DE and DM. In contrast, in the fully self-consistent code, $\beta(a,k)$ was chosen as a two-dimensional function $a$ and $k$, which is calculated by a modified code for anisotropies in the microwave background (CAMB). We assume that the DE perturbation is only effective at large scales. Thus, at small scales, gravity follows the usual Poisson equation. Therefore, only modifying the particle-mesh gravity solver for gravity at long wavelengths is accurate enough. The above four modifications in the N-body simulation have been considered upon improving the code Gadget2 \cite{gadget2}. The new code was named ME-Gadget.

%\subsection{Initial Condition}
In the simulation the capacity constrained Voronoi tessellation (CCVT) method was employed to produce a uniform and isotropic particle distribution and generate pre-initial conditions \cite{Liao2018}. In comparison to the gravitational equilibrium state (glass \cite{White1996}), the CCVT configuration is a geometrical equilibrium state, and is a more natural choice for models that include forces other than pure gravity. The open source code 2LPTic \cite{Crocce2006} is modified to read $H(a), m(a)$ generated by CAMB and is used to generate the initial condition for  simulations.  The second-order Lagrangian Perturbation Theory is applied following the equation describing the position displacement,
\begin{equation}
\mathbf{x(q)}=\mathbf{q}+\nabla_q\Psi^{(1)}+\nabla_q\Psi^{(2)},
\end{equation}
where $\Psi^{(1)}$ and $\Psi^{(2)}$ are first and second order displacement field, respectively.
The velocity displacement is given by
\begin{equation}
\mathbf{v(q)}=f_1H\nabla_q\Psi^{(1)}+f_2H\nabla_q\Psi^{(2)}\quad ,
\end{equation}
where 
\begin{equation}\label{f1f2}
f_{1,2}=\dfrac{dln(D_{1,2})}{dlna}\qquad .
\end{equation}
$\Omega_m$ and $H$ in interacting DE models are different from those in the $\Lambda$CDM model, where $f_1\approx\Omega_m^{3/5}$ and $f_2\approx 2\Omega_m^{4/7}$ \cite{Scoccimarro1998,Crocce2006}. In the modified 2LPTic code $f_1$, $f_2$, $\Omega_m$, and $H$ can be read at arbitrary redshifts as calculated by the modified CAMB for the interacting DE models \cite{Costa:2017}. However, at redshift $z=49$ and above the IDE models behave similarly to the standard model with $f_1\approx\Omega_m^{3/5}$ and $f_2\approx 2\Omega_m^{4/7}$.  Values of $f_{1,2}$ calculated from the modified CAMB can be used in the simulations.

\subsection{Nonlinear structure formation in interacting DE models }

\begin{figure}[thbp]
    \centering
    \includegraphics[width=0.48\linewidth]{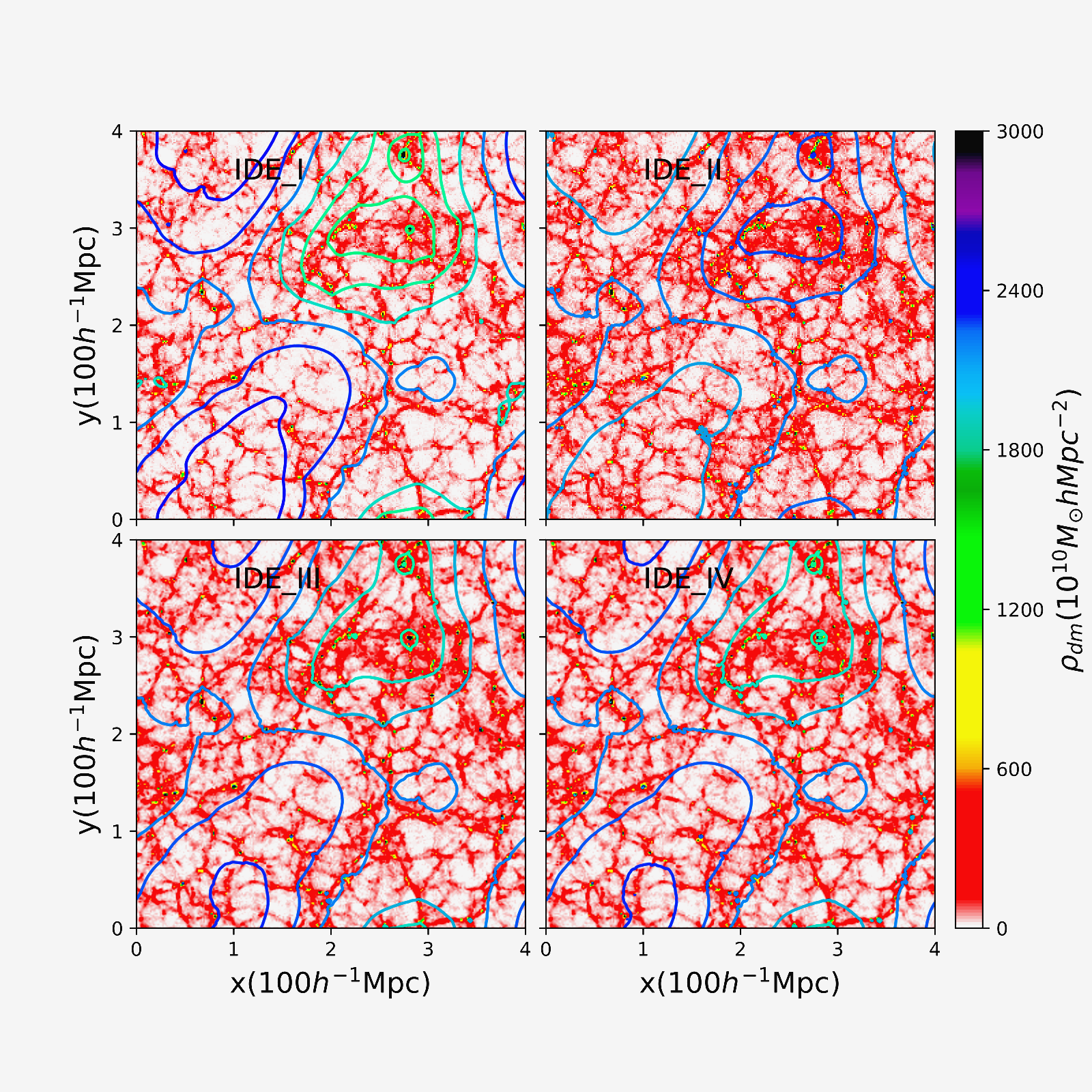}
    \includegraphics[width=0.48\linewidth]{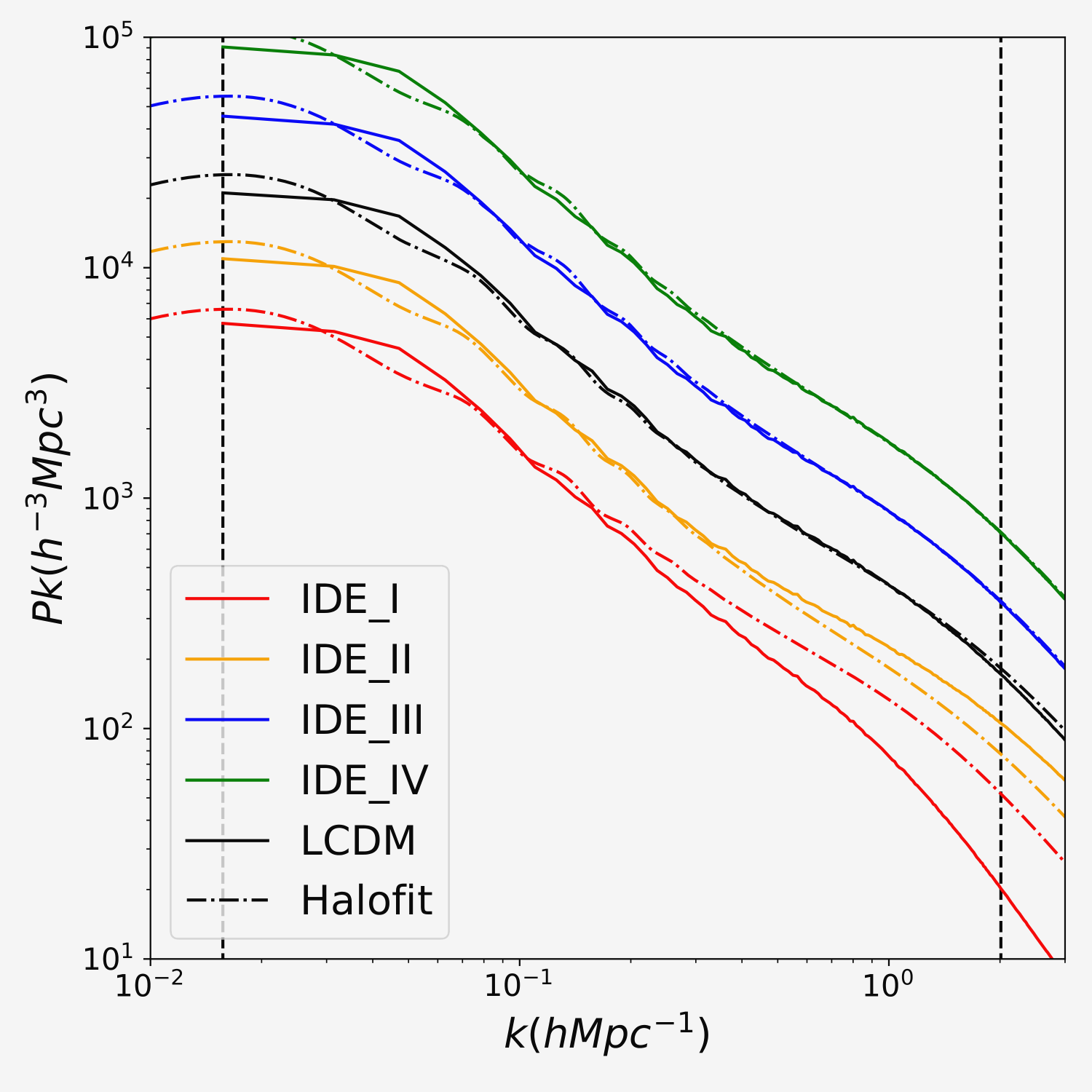}
    \caption{Left: The dark matter density distributions are shown in reddish colors and the dark energy distributions are shown in the contours. The colors of the contour represent the DE density, blue (green) contours represent low (high) DE density. The DE perturbation is only effective on large scales. Here we listed results for different interaction models. Right: The halofit approximation is valid for some specific weak interaction models between dark sectors. For general DE-DM interaction or strong coupling strength, we need fully nonlinear simulation to replace the halofit approximation.}
    \label{fig:X1}
\end{figure}

Running the ME-Gadget code, the fully self-consistent simulation can be performed with arbitrary inputs of the Hubble diagram, simulation particle mass, velocity dependent acceleration and DE perturbation to study  the non-linear structure formation in IDE models.  In the code DE perturbations at large scales, reasonable pre-initial conditions, initial matter power spectrum generated from CAMB code and  consistent second-order Lagrangian Perturbation Theory are fully considered.  Some results from the simulations are:  
\begin{itemize}
\item When there is an energy flow from DM into DE, the structure formation will be suppressed, and \textit{vice-versa}. The effect of the interaction between DM and DE on the non-linear evolution is non-trivial and it can be seen in the evolution of large scale structures. See for example Fig. \ref{fig:X1} Left.

\item The halofit approximation is valid only when the coupling between dark sectors is weak enough. If the interaction parameter is large, the halofit result is no longer appropriate and the fully nonlinear simulation is required. Please see Fig. \ref{fig:X1} Right.

\item DE perturbations grow together with DM density perturbations, but at much larger scales of $\sim100h^{-1}$Mpc. The growth of DE perturbations depends on the DE equation of state and the interaction parameters. 

\item For some interacting models, the results from the simulation on the
formation of nonlinear structures at $z=0$
are significantly different than those of the $\Lambda$CDM model.  This indicates that low redshift observations can be a powerful tool to 
constrain IDE models. Thus, Radio observations as provided by the BINGO/ABDUS project may be important tools \cite{Abdalla:2020,Abdalla2021a}
\end{itemize}

The ME-Gadget code has been used in multiple simulations with larger box sizes, higher mass resolutions, and larger parameter space to build up emulators of the observations. The simulations have been used to put further constraints on IDE models using data on large scale structures at low redshifts. Simulations that include IDE rooted on field theory are needed, to uncover the physics on the interaction within the dark sector through nonlinear structure formation in the universe \cite{Costa:2014pba}. 

\section{Prospects for future observations}
%\subsection{Future directions in understanding the interaction between dark sectors}
%\subsection{Prospects for future experiments}

Currently new facilities are being proposed, designed, constructed or
becoming operational that will provide new and more accurate cosmological data.
Forthcoming instruments
like the Euclid and Roman satellites will expand the area and depth
of current surveys. Further,
Stage V Spectroscopic Facilities (Spec-S5) carried out on a telescope with large aperture
and a wide field of view could measure the growth of structure at redshifts $2<z<5$
\cite{Annis:2022}.
Data from galaxy surveys like those of Dark Energy Survey Instrument (DESI,
\cite{Des:2023,Guy:2023})
or the Javalambre Physics of the Accelerating Universe Astrophysical Survey
(J-PAS, \cite{Jpas:2023}) are providing data on weak lensing, galaxy
clustering and growth of structure. DESI is currently conducting a survey of
$\sim 40$ million galaxies and quasars. It will measure the BAO peak position
with enough accuracy to determine angular diameter distances with
accuracy better than 0.3\% for $z<1.1$, 0.4\% for $1.1 < z < 1.9$, and 1\% at higher
redshifts. Satellites like Euclid and Roman will further improve these accuracies.

The Euclid satellite launched on July 2023 (\cite{Euclid:2023})
will measure $\sim 1.5$ billion galaxy shapes and photometric redshifts
and $\sim 35$ million redshifts with, respectively, $0.05(1+z)$ and $0.001(1+z)$ accuracy on 1/3 of the sky. The data on galaxy clustering,
gravitational lensing, BAOs
and RSDs will determine the expansion rate of the Universe
and the growth of cosmic structure with unprecedented accuracy, offering
probes into the nature of DM and DE. The Roman satellite has similar goals
but will carry them on a smaller fraction of the sky with larger depth.
Data from these satellites will allow to map Cosmic Infrared Background fluctuation
from source subtracted images \cite{Kashlinsky:2018} from where it will be possible
to recover the BAO signature at redshifts $z\ge 10$
\cite{Kashlinsky:2015}. These data will test the cosmological model at
redshifts not explored previously.

CMB facilities in operation
like the Simons Observatory (\cite{Simons:2023}) or planned like Stage IV (\cite{StageIV:2023}),
LiteBird (\cite{LiteBird:2023}), Probe of Inflation and Cosmic Origins (PICO, see \cite{Hanany:2019})
will provide independent observations that can be cross-checked to further
constrain models. Observations of the neutral Hydrogen 21cm line will probe
the epoch of reionization and the expansion rate, growth rate, and angular diameter distance
at those epochs (\cite{Bull:2015,Minoda:2023})

Most of the constraints on interacting DE models have come from large-scale and high-redshift observations, including cosmic expansion history, CMB, and large scale structures etc., see \cite{Wang:2016} and references therein, and some new results are summarized in  Sec.3. These observations can put tight constraints on interacting DE models when the interaction kernels depend on the DM density {\it like
those of type $Q_1$ (eq.~\ref{eq:background1}).}
For the interacting kernels proportional to the DE density, {\it like kernel $Q_2$,} since the DE becomes important only at the late time universe, it is expected that such interaction becomes significant at low redshifts, which has remarkable impact on the nonlinear structure formation.  New satellites like Euclid or Roman, CMB Stage IV instruments will test the evolution of
the cosmological background and first order density perturbations. Better prospects to constrain the interaction will come
from testing the model predictions in the nonlinear regime.

Cosmological N-body simulation is an essential tool to understand the nonlinear structure formation process. Employing the fully self-consistent numerical pipeline, the interacting DE model effect in the nonlinear structure formation was examined, see details summarized in Sec.4. It was found that at low redshifts some interacting DE models can be constrained tightly using the structure formation information.  To further unveil the nature of the interaction between DE and DM, we require complementary probes with some new physics.  In this Chapter we will discuss some prospects for future observations to disclose the interaction between dark sectors. We will emphasize on some new multi-messengers probes, including observations of the 21cm neutral hydrogen line and gravitational wave observations. The BINGO\footnote{Baryon acoustic oscillations from Integrated Neutral Gas
Observations (BINGO), \cite{Bingo:2023}} radio telescope is a unique instrument and the only radio telescope in South America designed to survey the cosmic large-scale structure using 21cm intensity mapping (IM). Measuring the coupling between DE and DM is one of its main scietific goals. We will also discuss how gravitational waves from mergers of solar and supermassive binary black holes as standard sirens can be used to further constrain the interaction between dark sectors in Cosmology.

\subsection{21cm cosmology}

Hydrogen is the most common baryon in our Universe. Neutral hydrogen (HI) has two almost degenerate ground states, split by the hyper-fine structure. The energy transition from the spin-1 state  to the spin-0 ground state can release 21cm radiation. The decay rate has an extremely small transition rate, implying a faint line. Observing the 21cm line via IM technique is one of the finest and most effective and reliable ways to map the large scale structure of our Universe. 

The BINGO radio telescope is specifically designed to study the properties of DE during the late-time cosmic acceleration by using IM. BINGO focuses on the redshift range $0.127 \le z \le 0.449$. It aims at providing new insights on physical properties of the interaction between DE and DM. The project has been described in the literature. The theoretical and instrumental aspects  are given in \cite{Abdalla2021a, Wuensche2021}, the optical design in \cite{Abdalla2021b}, the mission simulation in \cite{Liccardo2021}, further steps in component separation and bispectrum analysis in \cite{Fornazier2021}, a mock description in \cite{Zhang2021} and cosmological forecasting in \cite{Costa2021}. 

In the following we explain how the interaction between dark sectors can influence the 21cm brightness temperature fluctuation and its angular power spectrum. 

%%%%%%%%%%%%%%%%%%%%%%%%%%%%%%%%%%%%%%%%%%%%%%%%%%
%\subsection{HI Power Spectra}\label{sec.HI_theory}
The 21cm line originates from the transition between the hyperfine levels in the ground state of neutral hydrogen atoms, whose frequency in the rest frame is $\nu = 1420$\,MHz. The brightness temperature of the redshifted 21cm signal is of great interest to cosmology since the distribution of HI constitutes a good tracer of the large-scale structures in our Universe. The background brightness temperature is given by
\begin{align}
\bar{T}_{\mathrm{b}}(z) &= \frac{3}{32\pi} \frac{(h_{\mathrm{p}}c)^3\bar{n}_{\mathrm{HI}}A_{10}}{k_{\mathrm{B}}E_{21}^2(1+z)H(z)} \\
&= 0.188h \Omega_{\mathrm{HI}}(z) \frac{(1+z)^2}{E(z)} \mathrm{K},
\label{eq:Tb}
\end{align}
Now we focus on the perturbation of the brightness temperature to the linear order. Following~\cite{Hall2012}, in the conformal Newtonian gauge
\begin{equation}
\mathrm{d}s^2 = a^2(\eta)\left[(1+2\Psi)\mathrm{d}\eta^2 - (1-2\Phi)\delta_{ij}\mathrm{d}x^i\mathrm{d}x^j \right],
\label{eq:metric}
\end{equation}
with $\Psi$ and $\Phi$ spacetime-dependent gravitational potentials, the bulk velocity of HI tracing the total matter velocity reads $\mathbf{v}_\mathrm{m} \equiv \frac{\rho_{c}\mathbf{v}_{c} + \rho_{b}\mathbf{v}_{b}}{\rho_{c} + \rho_{b}}$ (the subscript $b$ here refers to baryon) and the corresponding Euler equation can be written as
\begin{equation}
\dot{\mathbf{v}} + \mathcal{H} \mathbf{v} + \mathbf{\nabla} \Psi = - \mathbf{v} \frac{ aQ}{\rho_{\mathrm{m}}} \,,
\label{eq:EulerEq}
\end{equation}
where $\rho_{\mathrm{m}}$ is the energy density for the total matter and the interaction between dark sectors reflects in the new term, $- \mathbf{v} \frac{ aQ}{\rho_{\mathrm{m}}}$. %Linfeng: I try to stress the new term we derived out here.
The perturbed brightness temperature $\Delta_{T_{\rm b}}$ including interaction between dark sectors  is given by
\begin{align}
\Delta_{T_{\rm b}}(z,\hat{\mathbf{n}}) & = \delta_n - \frac{1}{\mathcal{H}}\hat{\mathbf{n}} \cdot (\hat{\mathbf{n}} \cdot \mathbf{\nabla} \mathbf{v}) + \left(\frac{\rm d \ln (a^3 \bar{n}_{\text{HI}})}{\rm d \eta}-
\frac{\dot{\mathcal{H}}}{\mathcal{H}} - 2\mathcal{H} \right)\delta\eta \nonumber \\
&+ \frac{1}{\mathcal{H}}\dot{\Phi} + \Psi - \frac{1}{\mathcal{H}} \hat{\mathbf{n}} \cdot \mathbf{v} \frac{aQ}{\rho_{\mathrm{m}}},
\label{eq:perturb}
\end{align}
where $\delta_n$ is defined by $n_{\text{HI}} = \bar{n}_{\text{HI}}(1+\delta_n)$ and $\delta\eta$ is the perturbation of the conformal time $\eta$ at redshift $z$. We assume the large-scale clustering of HI gas follows the matter distribution, through some bias, and keep the conventional assumption that the bias is scale-independent. During the period of matter domination, where the comoving gauge coincides with the synchronous gauge, we can write $\delta_n$ in the Fourier space as~\cite{Hall2012}
\begin{equation}
\delta_n = b_{\rm HI}\delta_\mathrm{m}^{\text{syn}} +  \left( \frac{\rm d \ln (a^3
	\bar{n}_{\text{HI}})}{\rm d \eta} - 3 \mathcal{H}\right)\frac{v_\mathrm{m}}{k} ,
\label{eq.bias}
\end{equation}
where $k$ is the Fourier space wave vector, $v_{\mathrm{m}}$ is the Newtonian-gauge total matter velocity with $\mathbf{v} = - k^{-1} \mathbf{\nabla} v_{\mathrm{m}}$, $\delta_\mathrm{m}^{\text{syn}}$ is the total matter overdensity in the synchronous gauge and $b_{\rm HI}$ is the scale-independent bias.

In order to obtain the angular power spectrum of 21-cm line at a fixed redshift, we expand $\Delta_{T_{\rm b}}$ in spherical harmonics
\begin{equation}
\Delta_{T_{\rm b}}(z,\hat{\mathbf{n}})=\sum_{\ell m}\Delta_{T_{\rm b},\ell m}(z)Y_{\ell m}(\hat{\mathbf{n}}),
\end{equation} 
and express these perturbation coefficients $\Delta_{T_{\rm b},\ell m}(z)$ with the Fourier transform of temperature fluctuations, such that
\begin{equation}
\Delta_{T_{\rm b},\ell m}(z) = 4\pi i^l \int \frac{\rm d^3 \mathbf{k}}{(2\pi)^{3/2}}
\Delta_{T_{\rm b},\ell}(\mathbf{k},z) Y_{\ell m}^*(\hat{\mathbf{k}}).
\end{equation}
Following Eq.~(\ref{eq:perturb}), the $\ell$th multipole moment of $\Delta_{T_{\rm b}}$ reads
%\begin{align}
%\Delta_{T_b,l}(\mathbf{k},z)&=&\delta_n\, j_l(k \chi)+ \frac{kv}{\mathcal{H}}j_l{}''(k\chi)+\left(\frac{1}{\mathcal{H}}\dot{\Phi} + \Psi\right)j_l(k \chi)\\
%&&-\left(\frac{1}{\mathcal{H}}\frac{\rm d \ln (a^3 \bar{n}_{\text{HI}})}{\rm d \eta}-
%\frac{\dot{\mathcal{H}}}{\mathcal{H}^2} - 2 \right)\left[\Psi\, j_l(k \chi)+v\, j_l{}'(k \chi) +\int_0^\chi (\dot{\Psi}+\dot{\Phi})j_l(k \chi')d\chi' \right],\nonumber
%\end{align}
\begin{align}
\Delta_{T_{\rm b},\ell}(\mathbf{k},z) & = \delta_n\, j_{\ell}(k \chi)+ \frac{kv}{\mathcal{H}}j_{\ell}{}''(k\chi)+\left(\frac{1}{\mathcal{H}}\dot{\Phi} + \Psi\right)j_{\ell}(k \chi) \nonumber \\
& -\left(\frac{1}{\mathcal{H}}\frac{\rm d \ln (a^3 \bar{n}_{\text{HI}})}{\rm d \eta}-
\frac{\dot{\mathcal{H}}}{\mathcal{H}^2} - 2 \right) \left[\Psi\, j_{\ell}(k \chi) \right. \nonumber \\ 
& \left. +v\, j_{\ell}{}'(k \chi) +\int_0^\chi (\dot{\Psi}+\dot{\Phi})j_{\ell}(k \chi')d\chi' \right] \nonumber \\
& + \frac{1}{\mathcal{H}} v\, j_{\ell}{}'(k \chi) \frac{aQ}{\rho_{\mathrm{m}}}, 
\label{eq:perturb2}
\end{align}
where $\chi$ is the comoving distance to redshift $z$ and $j_{\ell}(k \chi)$ is the spherical Bessel Function. A prime on $j_{\ell}(k \chi)$ refers to a derivative with respect to the argument $k \chi$. Each term in Eq.~(\ref{eq:perturb2}) has its own physical meaning: $\delta_n$, in the first term, is the density fluctuation; the second term represents the effect of RSD; within the third term, $\dot{\Phi}/\mathcal{H}$ originates from the part of the ISW effect that is not cancelled by the Euler equation, whereas $\Psi$ arises from increments in redshift from radial distances in the gas frame. The physical meaning of those in the square brackets are very similar to the CMB contributions. The first, second and third terms correspond to the contributions from the usual SW effect, Doppler shift and ISW effect, respectively, from the perturbed time of the observed redshift. They are multiplied by a factor basically characterizing the time derivative of $\bar{T}_{\mathrm{b}}$ (i.e., $d\bar{T}_{\mathrm{b}}/d\eta$). The final term $\propto aQ$ is introduced by the interaction between the dark sectors.

We then integrate $\Delta_{T_{\rm b},\ell}(\mathbf{k},z)$ over a redshift (or frequency) normalized window function $W(z)$ as
\begin{equation}
\Delta_{T_{\rm b},\ell}^W(\mathbf{k}) = \int_0^\infty {\rm d} z W(z)
\Delta_{T_{\rm b},\ell}(\mathbf{k},z).
\label{eq:window}
\end{equation}
We assume a rectangular window function centered at redshift $z$ with a redshift bin width $\Delta z$ given by
\begin{equation}
W(z) = \begin{cases}
\frac{1}{\Delta z}, & z-\frac{\Delta z}{2}\leq z \leq z+\frac{\Delta z}{2}\,,\\
0, & \text{otherwise}\,.
\end{cases}
\end{equation}
Then the angular-cross spectrum of $\Delta_{T_{\rm b},\ell}$ between redshift windows can be calculated via
\begin{equation}
C_{\ell}^{WW'} = 4\pi\int {\rm d} \ln k \, {\cal P}_{\cal R}(k) \Delta_{T_{\rm b},\ell}^W(k) \Delta_{T_{\rm b},\ell}^{W'}(k).
\label{eq:Cl}
\end{equation}
${\cal P}_{\cal R}(k)$ is the dimensionless power spectrum of the
primordial curvature perturbation $\cal R$ and we define $\Delta_{T_{\rm b},\ell}^W(k) \equiv \Delta_{T_{\rm b},\ell}^W(\mathbf{k})/\cal R(\mathbf{k})$.

In Eq.~(\ref{eq:perturb2}) each term contains the information of the interaction between DE and DM. In the $\Lambda$CDM model, the auto-spectra for each term in ~(\ref{eq:perturb2}) with a channel bandwidth of $8.75\,\mathrm{MHz}$ at $z=0.28$, parameterized by the {\it Planck} 2018 best-fit values were listed in \cite{Xiao2019} and it was shown that  the density fluctuation and RSD term are the two leading contributions across the whole multipole range.  Especially at $\ell \sim 400$ the total signal is greatly dominated by the $\delta_n$ term. It would be interesting to see the influence of the interaction between DE and DM on each term in (\ref{eq:perturb2}) and compare with results in the standard cosmological model. Besides the auto-spectra, the total 21-cm signal also encompasses the contributions from cross-correlations of each two terms in (\ref{eq:perturb2}), which is in line with the case of real observation. Whether it can serve as a  smoking gun to disclose the interaction between dark sectors is worth careful investigation. As a novel approach, HI IM is of significant importance in revealing the interaction between dark sectors. 

BINGO will help understanding the interacting DE models. In Fig.3 we show that BINGO constraints on the IDE models are complementary to the Planck data constraints, which can help to further constrain the models. However astrophysical contamination is the main obstacle in extracting real data. Efficient foreground removal technique to separate the real HI IM signal from contamination is crucial to the success of HI IM probes. The systematic effect, primarily related to the instrument, prevents obtaining the clean signal at small scales, which needs more careful treatments.  After estimating the noise level, \cite{Xiao:2021} indicates a bright prospect for HI IM surveys in constraining interacting DE models. More discussions on noise removal can be found in BINGO series papers \cite{Abdalla2021a, Abdalla2021b, Wuensche2021, Liccardo2021, Fornazier2021, Zhang2021, Costa2021}.

\subsection{Effect of neutrinos}

Solar and atmospheric experiments suggested that neutrinos can have nonzero mass, however such experiments cannot pin down the absolute mass scale of neutrinos, instead they can only place a lower limit on the effective neutrino mass. In cosmology, neutrinos can play a very crucial role in the dynamics of our Universe. Massive neutrinos can become non-relativistic either before or after recombination \cite{Komatsu2011}, leaving an impact on the first acoustic peak in CMB temperature angular power spectrum or altering the matter-radiation equality, depending on their mass scale. They can also suppress the matter  power spectrum on small scales \cite{Hu1998,Yang62}.  Thus cosmological observations can put tight constraints on the sum of the neutrino masses. 
Assuming an extended base-$\Lambda$CDM cosmology with $3$ degenerate massive neutrinos, the most recent measurements from the Planck satellite on CMB in combination with BAO detections give an upper limit for the sum of the neutrino mass as $\sum m_\nu <0.12$ev ($95\%$ C.L.), on top of a benchmark $N_{\rm eff}$ value of $3.046$ \footnote{The latest effective number of neutrinos has been updated to $N_{\rm eff}  = 3.044$. See \cite{Froustey:2020mcq, Bennett:2020zkv} for detail.} referred by the standard model \cite{Planck_2018_6:2020}. A new degree of freedom in $N_{\rm eff}$ will incur almost invisible change to the $\Sigma m_{\nu}$ constraint, where the joint $95\%$ C.L. constraint stands on $N_{\rm eff} = 2.96^{+0.34}_{-0.33}$, derived by the same database of Planck 2018 TTTEEE + lowE + lensing + BAO. Certainly the total neutrino mass limit varies with the neutrino physics. A very recent work \cite{DiValentino:2021hoh} reported the most costraining bound, limited in the normal ordering case, to be $\Sigma m_{\nu} < 0.09$eV.
%Assuming the standard $\Lambda$CDM model, the most recent measurements from the Planck satellite on CMB in combination with the BAO and other low redshift observations give an upper limit for the sum of the neutrino mass as $\sum m_\nu <0.23$ev \cite{Xiaoask} and the effective number of neutrino species as $N_{eff}=3.04$. Including $f\sigma_8$ datasets, the upper limit on the sum of the neutrino mass can be constrained even lower. 

Cosmological constraints on neutrino properties in frameworks of different DE models have been investigated.  Massive neutrinos have also been studied in $f(R)$ gravity.  An incomplete list of past works can be found in \cite{Yang62,Yang63,Yang64,Yang65,Yang67,Yang68,Yang69,Yang70,Yang71,Yang72,Yang73,Yang74,Yang75,Yang76} and \cite{Kumar22,Kumar23,He2013}. Given the motivation and the recent indication in favor of an interaction between dark sectors, there have been a lot of attempts in exploring the interacting dark sector scenario with massive neutrinos by using the latest observational data \cite{Feng:2020a,Yang79,Yang80,Yang81,Yang82,Yang83,Yang84,Yang85,Kumar:2016}. It is interesting to discuss how the interaction between DE and DM affects the cosmological weighing of neutrinos and how massive neutrinos can in turn influence the dark interaction. It was shown that the presence of a sterile neutrino can provide new indication of interaction in dark sectors \cite{Kumar:2016}. Further evidences of interaction between DE and DM serving as an alternative scenario to explain the observable universe were given in presence of neutrino properties \cite{Yang:2020a, Guo:2017}.

In addition to available observations including CMB, BAO, $H_0$, $f\sigma_8$ etc., in studying the $\Lambda$CDM model, it was found that the combination of CMB data with measurements of the HI power spectrum can improve the constraint of the total mass of neutrino. Combing the covariance matrix from the Planck data with 21cm Fischler analysis, $\sum m_\nu <0.14$ev (95\% CL, BINGO+Planck)\cite{Costa2021}.  

It will be of great interest to extend the discussion to the framework of interacting DE models and constrain total neutrino mass and number of relativistic species combing with the 21cm observations. Besides the linear 21cm spectrum, considering possible significant deviation from the massless neutrino result may show up on small scales, it will be advantageous to study neutrino physics in the nonlinear 21cm signal using N-body simulation. More studies in these directions are called for. 

\begin{figure}[thbp]
    \centering
    \includegraphics[width=0.48\linewidth]{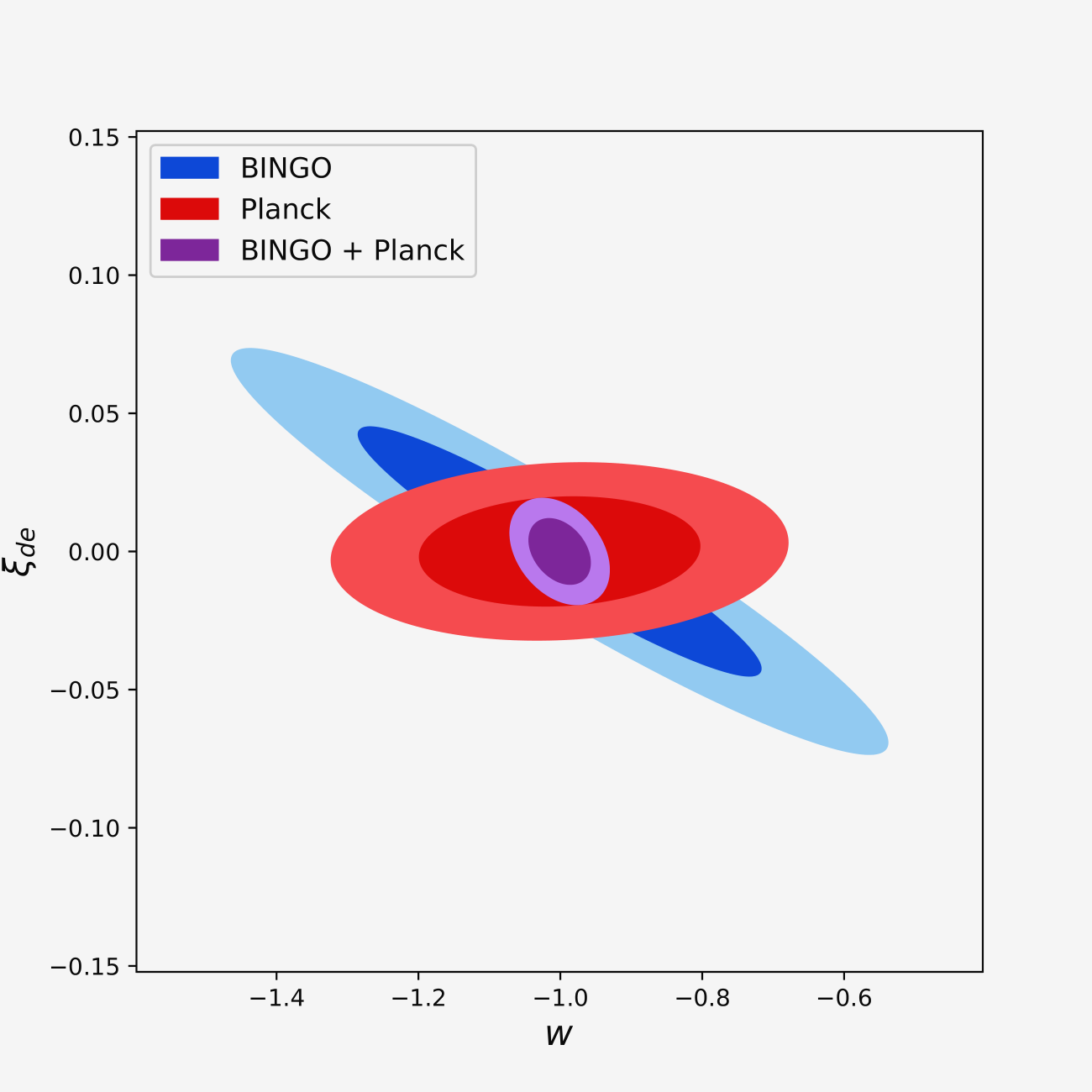}
    \includegraphics[width=0.48\linewidth]{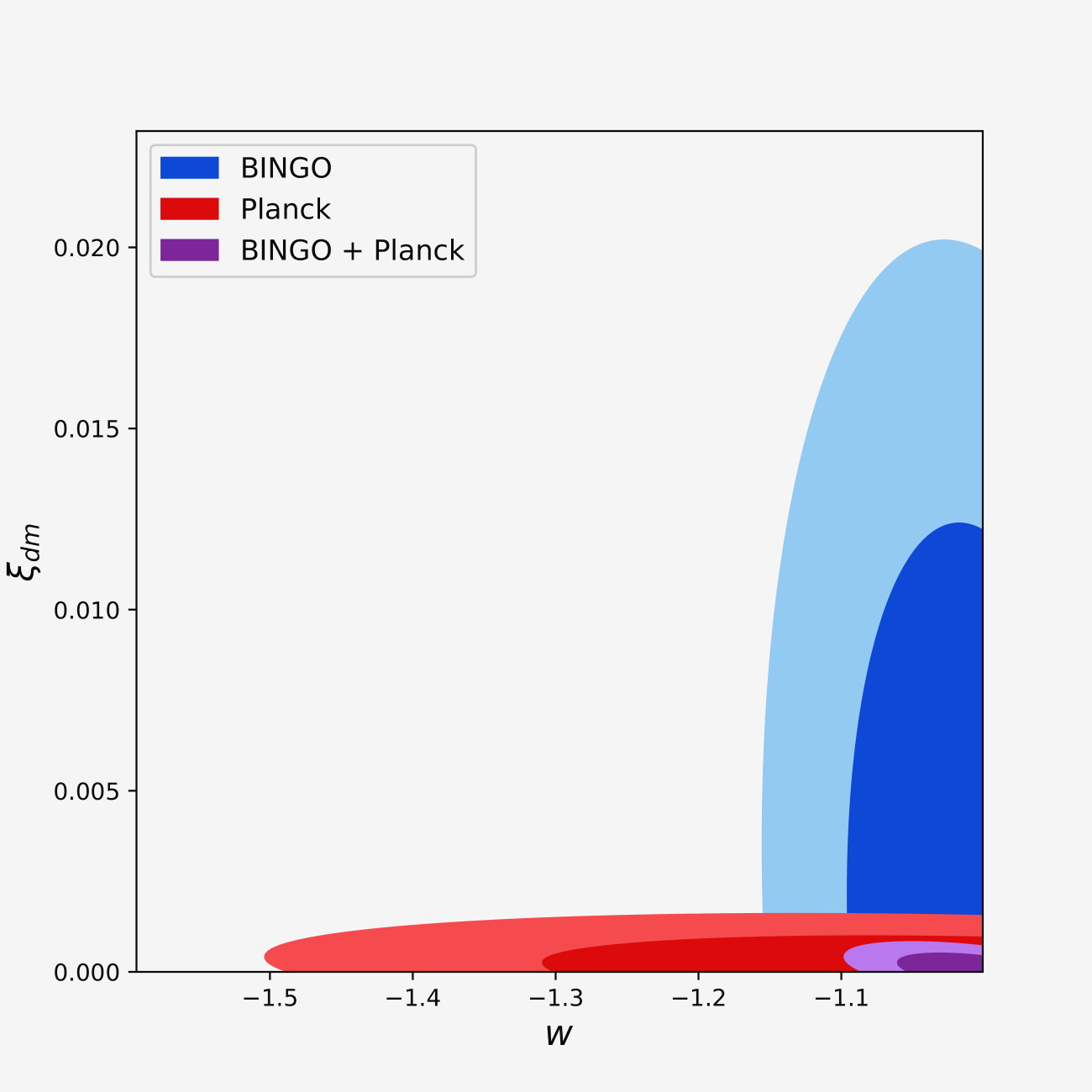}
    \caption{We present here constraints on interaction models using: BINGO and Planck.}
    \label{fig:X3}
\end{figure}

\subsection{Fast Radio Bursts as cosmological probes}

Fast Radio Bursts (FRBs) are millisecond emissions of radio signals originated at cosmological distances. Since the first discovery in 2007, hundreds of FRBs have been published to date. FRBs are expected to be useful probes of cosmology.  The dispersion measures of FRBs, combined with the redshifts of their host galaxies have potential to provide a good map of the baryon distribution, magnetic fields and turbulence in the intergalatic medium (IGM), helium reionization, etc. The dispersion measure of FRBs is a convolution of the cosmic distance and can be used as a new cosmological distance measure to constrain the DE equation of state, the Hubble parameter and the reionization history. Recent reviews on FRBs progenitor theories and emission mechanisms can be found in \cite{Platts2018}\cite{Dai2021}\cite{Flynn2021}\cite{Zhang2022}. 

Radio waves in a plasma are dispersed, with waves with lower frequencies delayed with respect to waves with higher frequencies. The dispersion measure $\widehat{\mathcal{DM}}$ describes the degree of such a delay. The best-fit $\widehat{\mathcal{DM}}$ is obtained for each FRBs when it is discovered. 

The observed $\widehat{\mathcal{DM}}$ is usually split into multiple terms 
\begin{equation}
    {\widehat{\mathcal{DM}}= \widehat{\mathcal{DM}}_{\rm MW} + \widehat{\mathcal{DM}}_{halo} + \widehat{\mathcal{DM}}_{IGM}} + \frac{\widehat{\mathcal{DM}}_{host} + \widehat{\mathcal{DM}}_{src}}{1+ z},
\label{eq:DMterms}
\end{equation}
where $\widehat{\mathcal{DM}}_{\rm MW}$, $\widehat{\mathcal{DM}}_{halo}$, $\widehat{\mathcal{DM}}_{IGM}$, $\widehat{\mathcal{DM}}_{host}$, and $\widehat{\mathcal{DM}}_{src}$ are the contributions from the Milky Way, its halo, the IGM, the host galaxy, and the immediate environment of the source, respectively. 

The observed contributions from the last two components are smaller by a factor of $(1+z)$, where $z$ is the source redshift. The Milky Way (MW) term  can be obtained using MW electron density models. The extended MW halo contributions can also be estimated. The average value of intergalactic dispersion measure relates to cosmological models 
\begin{equation}
    \left< {\widehat{\mathcal{DM}}_{IGM}(z)} \right> = \frac{3 c H_0 \Omega_b f_{\rm IGM}}{8\pi G m_p} \int_0^z \frac{\chi(z')(1+z') dz'}{E(z')},
\label{eq:DM-z}
\end{equation}
where
\begin{equation}
    \chi(z) \simeq \frac{3}{4} \chi_{\rm e,H} (z) + \frac{1}{8} \chi_{\rm e,He} (z)
    \label{eq:chiz}
\end{equation}
noticing that the cosmological mass fractions of HI and He are $\sim 3/4$ and $\sim 1/4$, respectively, $\Omega_b$ is the energy density fraction of baryons, $f_{\rm IGM}$ is the fraction of baryons in the IGM, and $\chi_{\rm e,H}(z)$ and $\chi_{\rm e,He}(z)$ are the fractions of ionized electrons in hydrogen (HI) and helium (He), respectively, as a function of redshift. 

The theoretical predictions and the implications of the $\widehat{\mathcal{DM}}_{IGM}(z)$ relation have been a major driver of further FRB observations as astrophysical and cosmological probes.  It was argued that the dispersion measure could potentially offer complementarity with standard distance probes to constrain cosmological models \cite{Linder2019}. 

In \cite{Gao2014,Walter2018,Zhou2014,Dai252,Dai253},  meaningful constrains on DE was achieved with a large enough FRB samples, especially in combination with other cosmological probes such as CMB, BAO, SNIa, Gamma Ray Bursts and gravitational wave observations. In recent studies \cite{Dai253,Zhao2022} investigations were done on the capability of future FRB data for improving cosmological parameter estimation in dynamical and interacting DE models. 

Current surveys, such as Green Bank, CHIME and FAST telescopes etc., are playing important roles in detecting FRBs.  BINGO is the upcoming survey which will start to collect data in 2024 and has the potential to detect and localize FRBs through its main telescope, outriggers and phase arrays. As initially investigated in \cite{Abdalla2021a}, BINGO will be able to detect  $\sim 113$ FRBs per year and localize more than circa $ 50$ per year, using nine 6 m dishes as outriggers. Using a phased array of 6 m diameter, BINGO will localize 10 times more than the corresponding value for a set of nine 6 m-outriggers. Also,  BINGO will localize FRBs on higher redshifts using the phased array, up to $z\sim 1.3$. In the scenario where we have two 40 m-phased array stations, the number of detections and localizations almost doubles and the localizations will occur up to redshifts of $\sim 3.5$. In Fig.4, FRB detection rate (per year) for BINGO and outriggers are illustrated.
Further combining with the future mega-SKA project, it is expected that more FRB events can be detected, which will make the constraint of cosmological parameters feasible, including signatures of interactions between dark sectors.  Considering DM interacting with DE will influence the local structure formation, which in turn will affect  $\widehat{\mathcal{DM}}$ contributed by host galaxies and fluctuations of IGM, more careful investigations are needed to use FRBs to constrain interacting DE models. 

\begin{figure}[thbp]
    \centering
    \includegraphics[width=0.7\linewidth]{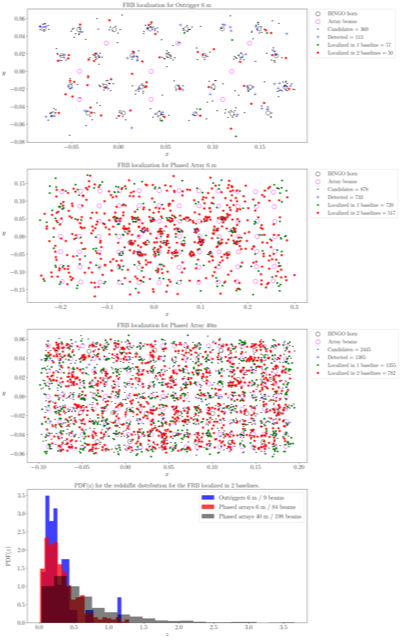}
    \caption{FRB detection rate (per year) for BINGO and outriggers. Top panel shows the detection and localization rate for BINGO main horns and a set of nine outriggers of 6 m diameter. The second panel assumes two phased array stations of 6 m diameter. The third panel assumes two Mantis-like stations as outriggers. The x and y axes are the altitude-azimuth coordinates. The bottom panel presents the redshift distribution of the locations in 2 base-lines for the different configurations.}
    \label{fig:X2}
\end{figure}

\subsection{Gravitational waves as standard sirens}

Mapping the expansion history of the universe through distance-redshift relation is a direct observational probe of cosmological parameters.  This relation was mainly learnt from observations of distant SNIa, which serve as standard candles that reveal their luminosity distances.  Systematic errors in such observations are of particular concern. Gravitational waves (GW) from massive binary black holes in-spirals are potential powerful standard sirens, which serve as a completely independent measurement of the distance-redshift relation \cite{Schutz, Schutz2002, Hughes2002}.  The binary black hole systems are relatively simple and well modeled, the GWs they generate determine the source's luminosity distance with high accuracy. 

For the inspiral of a compact binary system with component masses $m_1$ and $ m_2$, the two polarizations of GW signal  can be described as \cite{Colpi:2016fup}
\begin{subequations}
 \begin{align}
 h_+ (t) &=  \left(\frac{G \mathcal{M}_z}{c^2} \right)^{5/3} \left(\frac{\pi f(t)}{c} \right)^{2/3} \frac{2 (1+ \cos^2 \iota)}{D_L} \cos \big(\Psi (t, \mathcal{M}_z, \eta) \big),    \label{h_plus}    \\
 h_{\times} (t) &=  \left(\frac{G \mathcal{M}_z}{c^2} \right)^{5/3} \left(\frac{\pi f(t)}{c} \right)^{2/3} \frac{4 \cos \iota}{D_L} \sin \big(\Psi (t, \mathcal{M}_z, \eta) \big),    \label{h_cross}
 \end{align}
\end{subequations}
where $\mathcal{M}_z = (1+z)\mathcal{M} = (1+z)(m_1m_2)^{3/5}/(m_1+m_2)^{1/5}$ is the redshifted chirp mass (with a perfect degeneracy between the redshift and the physical mass), $\eta = m_1 m_2 / (m_1 + m_2)^2$ is the symmetric mass ratio, $\iota$ is the inclination angle of the binary orbital angular momentum relative in the line of sight, and $\Psi (t, \mathcal{M}_z, \eta)$ is the phase of the GW signal.
Notably, the overall amplitude is determined solely by the redshifted chirp mass, the inclination, and the luminosity distance.
By observing the GW phase evolution, the mass parameter $\mathcal{M}_z$ can be reliably estimated, and the inclination angle can be determined by observing the amplitude ratio of different polarizations. The luminosity distance $D_L$ influenced by cosmological parameters can be obtained from the GW data released in compact binary coalescence. 

The largest uncertainty in determining $D_L$ through GW arises from the pointing errors. GWs do not provide the redshift of the source. If redshifts of mergers can be inferred, one can use information from both $D_L$ and $z$ to constrain cosmological parameters. This makes the spiralling signal of compact binary systems desirable for cosmological studies. If an electromagnetic counterpart to a binary black hole GW event exists, this concern can be released \cite{Schutz, Schutz2002}. However if there is no electromagnetic counterpart, precise localizations of GW sources are needed for cosmological parameters constraints. Possible collaborations among different space detectors to improve GW source localization was discussed in \cite{Gong2021}.  Further carefully evaluating gravitational lensing and peculiar velocity, the intrinsic uncertainty of $D_L$ estimation can be reduced \cite{Zhu2022}.  

Employing the standard sirens from GWs released from supermassive binary black hole spiralling, constrains of cosmological parameters have been obtained, including models with interaction between DE and DM \cite{2211, Bachega:2020, Yang:2019b, Yang:2020b, 1903, 1901, 1803, Cai:2017, Caprini:2016, 1312}. It was found that the addition of the GWs data to other astronomical measurements significantly improves the constraints on parameters space. GW sirens can constrain the dark sector interaction at redshift ranges not reachable by usual supernovae datasets. It is expected that with a larger number of GW events, GW standard sirens can provide further tests to the interacting DE models.

\subsection{Forecasts for future experiments.}

Cosmic variance limited data on CMB temperature and polarization
anisotropies and BAOs will significantly
improved the equation of state parameter of the DE, although
further information would be needed to measure a non-zero coupling
\cite{Joseph:2023}.
The new observational facilities would add new and better information on
to dilucidate the capacity of future measurements of angular diameter
and luminosity distances, CMB ansisotropies, cosmic shear, matter growth
experiments, etc, to constrain phenomenological interacting dark energy models.
Dark energy starts to dominate the dynamics of the Universe at $z\sim 1$
and surveys like BINGO and SKA, sensitive to this redshift, will probe the
effect of the DM-DE interaction \cite{Xu:2018}.

Gravitational waves from  merging binary systems
together with observation of their electromagnetic part
will become standard sirens that, like astronomical
standard candles, will probe the expansion rate of the Universe
in the intermediate redshift range $1\le z\le 8$, which make
them ideal to test the transition from a matter dominated to
accelerated expansion in the presence of DM-DE interactions
\cite{Bachega:2020,Baral:2021,Cai:2017,Califano:2023,Caprini:2016}.
Adding GW data could detect the interaction at the 3$\sigma$ C.L. \cite{Yang:2019b},
improve the statistical significance of the null result \cite{Yang:2020b}
or reconstruct the coupling function \cite{Bonilla:2022}.

Forthcoming neutral hydrogen intensity mapping at 21cm from facilities
like BINGO, FAST, SKA1-MID and Tianlai, in combination with Planck data
would detect the strength of the interaction of kernels $Q_1,Q_2,Q_3$
(eq.~\ref{eq:kernels}) with amplitudes $\sim 10^{-3}$ \cite{Zhang:2021}.
In \cite{Xiao:2021} the Fisher matrix is computed for these future datasets
to forecast constraints on the IDE models.

Observational constrains from J-PAS \cite{Costa:2019,Figueruelo:2021,Salzano:2021}
will test the nature of dark energy and show if interacting models alleviate the
$S_8$ tension \cite{Beltran:2021b}.
Different observables will be more or less sensitive to DM-DE interaction
depending on the redshift. With improved data on redshift space distortions,
cosmic magnification and galaxy counts the imprint of
a DM-DE interaction can be probed out to $z\le 3$ \cite{Duniya:2022a,Duniya:2022b}
The forthcoming data on BAOs from Euclid and high precision CMB temperature
and polarization anisotropies could potentially exclude the null interaction
for kernels $Q_1$ and $Q_2$ (see eq.~\ref{eq:kernels}) \cite{Santos:2017}.

\section{Conclusions}

The existence of a hypothetical  interaction between dark matter and dark energy affects the formation and evolution of the large scale structure in the Universe. It is a degree of freedom that needs to be considered when analyzing cosmological observational data. Originally introduced to solve the coincidence problem, IDE model are currently analyzed in order to explain away the discrepancies between the predictions of the
$\Lambda$CDM model and observation. Lacking a theoretical motivation about the nature of the
DM-DE interaction, most models considered until now are phenomenological. The interaction kernel is postulated and the model predictions are tested against observations. Currently, as we have
shown in this review, the data on the evolution of the background cosmology and linear perturbations does not show a clear preference for models with interactions. Given the large variety of interacting kernel, data sets and statistical tools, it is difficult to draw a definitive conclusion. The observations of the dynamics of the background and linear evolution of density perturbations may  not  fully compatible among themselves since they could be affected by systematics. We have argued that non-linear effects are a more sensitive probe of the interaction but estimating those effects requires expensive computer simulations. New theoretical tools to understand how the interaction affects the non-linear regime are required to make further progress. On the observational side, new facilities like the Euclid and Roman satellites, 21cm observations by BINGO, data from gravitational waves, etc, promise more extensive and accurate data to test models further. A clear determination of an interaction kernel would be a result of fundamental importance since it would provide a direct insight into fundamental problems in physics today: the physical nature and properties of the dark components, dark matter and dark energy.

\end{document}